\documentclass[aps,pra,twocolumn,superscriptaddress,letterpaper]{revtex4-1}
\usepackage{graphicx}
\usepackage{amsmath}
\usepackage{amssymb}
\usepackage{esint}
\usepackage{verbatim}
\usepackage{soul}
\usepackage{hyperref}

\newcommand{\be}{\begin{equation}}
\newcommand{\ee}{\end{equation}}
\newcommand{\ba}{\begin{array}}
\newcommand{\ea}{\end{array}}
\newcommand{\bqa}{\begin{eqnarray}}
\newcommand{\eqa}{\end{eqnarray}}

\newcommand{\nbar}{\langle n \rangle}
\newcommand{\nbath}{n_{\text{b}}}
\newcommand{\nnaught}{n_{0}}
\newcommand{\nbathp}{n_{\text{p}}}
\newcommand{\Tbathp}{T_{\text{p}}}

\newcommand{\Tbathnaught}{T_{0}}

\newcommand{\Tf}{T_{\text{f}}}

\newcommand{\gammab}{\gamma_{\text{b}}}

\newcommand{\nmi}{n_{\text{i}}}
\newcommand{\nmf}{n_{\text{f}}}

\newcommand{\lambdac}{\lambda_{\text{c}}}
\newcommand{\ncav}{n_{\text{c}}}
\newcommand{\nwg}{n_{\text{wg}}}

\newcommand{\gzero}{g_{\text{0}}}
\newcommand{\kappai}{\kappa_{\text{i}}}
\newcommand{\kappae}{\kappa_{\text{e}}}
\newcommand{\gammap}{\gamma_{\text{p}}}
\newcommand{\gammanotO}{\gamma_{\text{0}}}
\newcommand{\gammaOM}{\gamma_{\text{OM}}}
\newcommand{\Qo}{Q_{\text{c}}}
\newcommand{\Qm}{Q_{\text{m}}}

\newcommand{\omegal}{\omega_{\text{l}}}

\newcommand{\omegas}{\omega_{\text{s}}}
\newcommand{\omegap}{\omega_{\text{p}}}
\newcommand{\omegac}{\omega_{\text{c}}}

\newcommand{\Ceff}{C_{\text{eff}}}
\newcommand{\Pin}{P_{\text{in}}}

\newcommand{\gammaSB}{\Gamma_\text{SB,0}}

\newcommand{\etaSPD}{\eta_\text{SPD}}

\newcommand{\omegam}{\omega_{\text{m}}}

\newcommand{\Toff}{\tau_\text{off}}
\newcommand{\Tpulse}{\tau_\text{pulse}}
\newcommand{\Tper}{\tau_\text{per}}

\newcommand{\Qoscat}{Q_\text{c,scat}}
\newcommand{\Pthm}{P_\text{th}}
\newcommand{\Qoi}{Q_\text{c,i}}
\newcommand{\gammatot}{\gamma}

\newcommand{\gammaphi}{\gamma_{\phi}}
\newcommand{\thermcond}{C_{\text{th}}}

\newcommand{\epsilononeD}{\epsilon_\text{1D}}
\newcommand{\epsilontwoD}{\epsilon_\text{2D}}
\newcommand{\PthoneD}{P_\text{th,1D}}
\newcommand{\PthtwoD}{P_\text{th,2D}}

\begin{document}

\title{Two-Dimensional Optomechanical Crystal Cavity with High Quantum Cooperativity}

\author{Hengjiang Ren}
\author{Matthew H. Matheny}
\author{Gregory S. MacCabe}
\author{Jie Luo}
\affiliation{Kavli Nanoscience Institute, California Institute of Technology, Pasadena, California 91125, USA}
\affiliation{Institute for Quantum Information and Matter and Thomas J. Watson, Sr., Laboratory of Applied Physics, California Institute of Technology, Pasadena, California 91125, USA}
\author{Hannes Pfeifer}
\altaffiliation[Current Address: ]{Institut f\"{u}r Angewandte Physik, Universit\"{a}t Bonn, Wegelerstraße 8, 53115 Bonn, Germany}
\affiliation{Max Planck Institute for the Science of Light, Staudtstrasse 2, 91058 Erlangen, Germany}
\author{Mohammad Mirhosseini}
\author{Oskar Painter}
\thanks{opainter@caltech.edu; http://copilot.caltech.edu}
\affiliation{Kavli Nanoscience Institute, California Institute of Technology, Pasadena, California 91125, USA}
\affiliation{Institute for Quantum Information and Matter and Thomas J. Watson, Sr., Laboratory of Applied Physics, California Institute of Technology, Pasadena, California 91125, USA}
\email{opainter@caltech.edu; }

\date{\today}

\begin{abstract} 

    Optomechanical systems offer new opportunities in quantum information processing and quantum sensing. Many solid-state quantum devices operate at millikelvin temperatures -- however, it has proven challenging to operate nanoscale optomechanical devices at these ultralow temperatures due to their limited thermal conductance and parasitic optical absorption. Here, we demonstrate a two-dimensional optomechanical crystal resonator capable of achieving large cooperativity $C$ and small effective bath occupancy $\nbath$, resulting in a quantum cooperativity $\Ceff\equiv C/\nbath \approx 1.3 > 1$ under continuous-wave optical driving.  This is realized using a two-dimensional phononic bandgap structure to host the optomechanical cavity, simultaneously isolating the acoustic mode of interest in the bandgap while allowing heat to be removed by phonon modes outside of the bandgap. This achievement paves the way for a variety of applications requiring quantum-coherent optomechanical interactions, such as transducers capable of bi-directional conversion of quantum states between microwave frequency superconducting quantum circuits and optical photons in a fiber optic network.

\end{abstract}

\maketitle


Recent advances in optomechanical systems, in which mechanical resonators are coupled to electromagnetic waveguides and cavities~\cite{Kippenberg2008,Aspelmeyer2014}, have led to a series of scientific and technical advances in areas such as precision sensing~\cite{Clerk2010,hanay2015inertial}, nonlinear optics~\cite{Hill2012,Safavi-Naeini2013b}, nonreciprocal devices~\cite{fang2017generalized,peterson2017demonstration,bernier2017nonreciprocal}, and topological wave phenomena~\cite{schmidt2015optomechanical,brendel2017pseudomagnetic}. In addition, such systems have been used to explore macroscopic quantum phenomena, from initial demonstrations of laser cooling of mechanical resonators into their quantum ground state~\cite{teufel2011sideband,Chan2011,Verhagen2012,peterson2016laser,rossi2018measurement}, to heralded preparation and entanglement of mechanical quantum states~\cite{riedinger2016non,hong2017hanbury,riedinger2018remote,Marinkovi2018}, generation of squeezed light~\cite{Safavi-Naeini2013b,Purdy2013b}, and coherent transduction between photons with different energies~\cite{Hill2012,Bochmann2013,Andrews2014,balram2016coherent,higginbotham2018harnessing,bienfait2019phonon}.

Optomechanical crystals (OMCs)~\cite{Eichenfield2009b}, where electromagnetic and elastic waves overlap within a lattice, are patterned structures that can be engineered to yield large radiation-pressure coupling between cavity photons and phonons. Previous work has realized one-dimensional (1D) silicon (Si) OMC cavities with extremely large vacuum optomechanical coupling rates ($\gzero \approx 1$~MHz)~\cite{Chan2012,matheny2018enhanced}, enabling a variety of applications in quantum optomechanics including the aforementioned ground-state cooling~\cite{Chan2011} and remote quantum entanglement of mechanical oscillators via an optical channel~\cite{riedinger2018remote}.  An application area of growing interest for OMCs is in hybrid quantum systems involving microwave-frequency superconducting quantum circuits~\cite{schoelkopf2008wiring,devoret2013superconducting}.  Owing to the large ratio ($\times 10^5$) of the speed of light to the speed of sound in most materials, OMCs operating at telecom-band optical frequency naturally couple strongly to similar wavelength microwave-frequency acoustic modes.  Recent experimental demonstrations of microwave-frequency phononic crystal cavities with ultralow dissipation~\cite{MacCabe2019} and strong-dispersive coupling to superconducting qubits~\cite{Arrangoiz_Arriola2019} indicate that there are potentially significant technical advantages in forming an integrated quantum electrodynamic and acoustodynamic circuit architecture for quantum information processing~\cite{pechal2018superconducting,hann2019hardware}.  In such an architecture, OMCs could provide a quantum interface between microwave-frequency logic circuits and optical quantum communication channels.          

A significant roadblock to further application of OMC cavities for quantum applications is the very weak, yet non-negligible parasitic optical absorption in current devices~\cite{Meenehan2015b,riedinger2016non,hong2017hanbury,riedinger2018remote,Marinkovi2018}. Optical absorption, thought to occur due to surface defect states~\cite{Stesmans1996,Borselli2007}, together with inefficient thermalization due to the 1D nature of Si nanobeam OMC cavities currently in use, can yield significant heating of the microwave-frequency acoustic mode of the device. At ultralow temperatures ($\lesssim 0.1$~K), where microwave-frequency systems can be reliably operated as quantum devices, optical absorption leads to rapid (sub-microsecond) heating of the acoustic cavity mode~\cite{Meenehan2015b}.  This has limited quantum optomechanical experiments to schemes with high optical power and short pulses~\cite{Meenehan2015b,riedinger2016non,hong2017hanbury,riedinger2018remote,Marinkovi2018,MacCabe2019} or very low continuous optical power~\cite{Meenehan2014,qiu2018motional}.

\begin{figure*}[tp]
\begin{center}
\includegraphics[width=2\columnwidth]{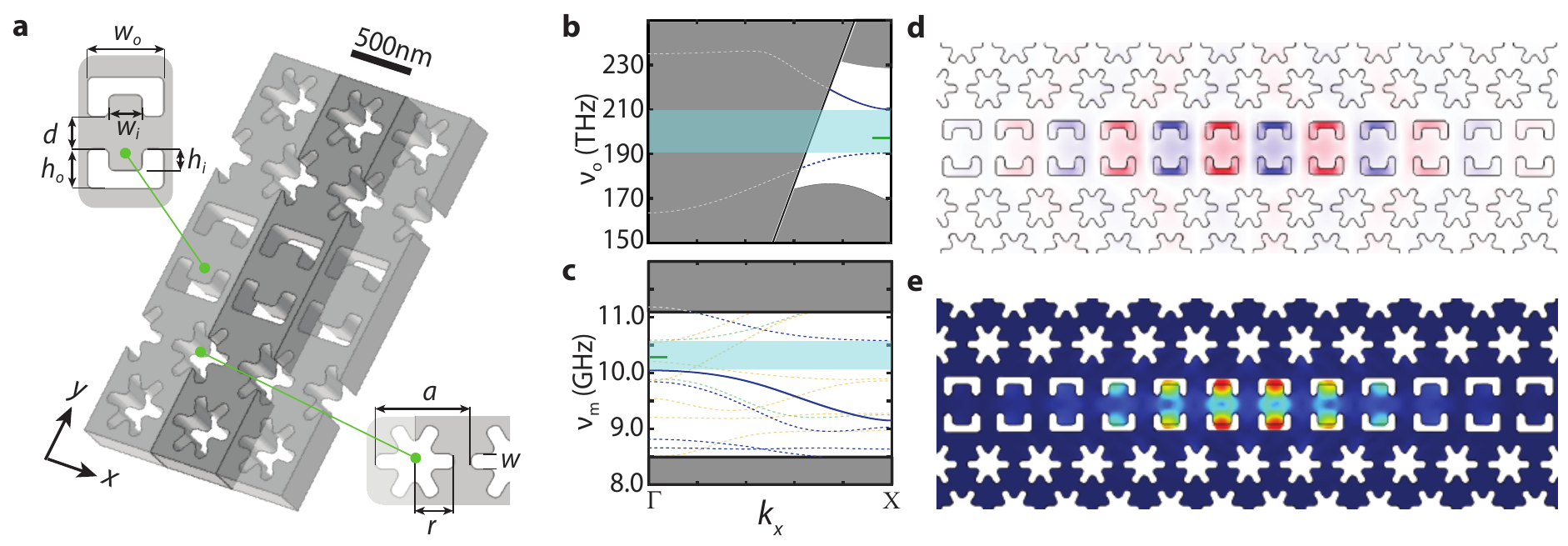}
\caption{\textbf{Quasi-2D OMC cavity design.} 
\textbf{a}, Unit cell schematic of a linear waveguide formed in the snowflake crystal. Guided modes of the waveguide propagate along the \textit{x}-axis. Insets: (left) `C' shape parameters, (right) snowflake parameters.
\textbf{b}, Photonic and \textbf{c}, phononic bandstructure of the linear waveguide.
The solid blue curves are waveguide bands of interest; dashed lines are other guided modes; shaded light blue regions are band gaps of interest; green tick mark indicates the cavity mode frequencies; gray regions denote the continua of propagating modes.  In the photonic bandstructure only modes of even vector parity about the center of the Si slab ($\sigma_z = +1$) are shown.  In the acoustic bandstructure green dashed curves are for $\sigma_\text{z} = +1$ and $\sigma_\text{y} = -1$ parity modes, and yellow dashed curves denote $\sigma_\text{z} = -1$ modes.
\textbf{d}, FEM-simulated mode profile ($E_\text{y}$ component of the electric field) of the fundamental optical cavity resonance at $\omegac/2\pi = 194$~THz, with red (blue) corresponding to positive (negative) field amplitude. 
\textbf{e}, Simulated displacement profile of the fundamental acoustic cavity resonance at $\omegac/2\pi = 10.27 \text{~GHz}$. The magnitude of displacement is represented by color (large displacement in red, zero displacement in blue).
}
\label{fig1}
\end{center}
\end{figure*}

The most relevant figure-of-merit for quantum optomechanical applications is the effective quantum cooperativity ($\Ceff\equiv C/\nbath$), corresponding to the standard photon-phonon cooperativity ($C$) divided by the Bose factor of the effective thermal bath ($\nbath$) coupled to the acoustic mode of the cavity~\cite{Hill2012,Andrews2014,Meenehan2015b}. In previous experiments with nanobeam OMC cavities at millikelvin temperatures, the quantum cooperativity was substantially degraded below unity (the relevant threshold for coherent photon-phonon interactions) due to the heating and damping caused by the optical-absorption-induced hot bath.  The heating of the acoustic cavity mode by the optically-generated hot bath can be mitigated through several different methods. The simplest approach in a low temperature environment is to couple the cavity more strongly to the surrounding cold bath of the chip, or, through addition of another cold bath as in experiments in a $^3$He buffer gas environment~\cite{shomroni2019optical,qiu2019high}. This method can be quite effective in decreasing the acoustic mode thermal occupancy in the presence of optical absorption; however, the effectiveness of the method relies on increasing the coupling to baths other than the optical channel, which necessarily decreases the overall photon-phonon quantum cooperativity.  

Here we employ a strategy that makes use of the frequency-dependent density of phonon states within a phononic bandgap structure to overcome this limitation. Using a two-dimensional (2D) OMC cavity~\cite{Safavi-Naeini2010,Gavartin2011,Safavi-Naeini2013} the thermal conductance between the hot bath and the cold environment is greatly increased due to the larger contact area of the 2D structure with the bath, while the acoustic mode of interest is kept isolated from the environment through the phononic bandgap of the structure. By keeping the intrinsic damping of the acoustic mode low, this method is a promising route to realizing $\Ceff > 1$.  Initial work in this direction, performed at room temperature, utilized snowflake-shaped holes in a Si membrane to create a quasi-2D OMC with substantially higher optical power handling capability, although with a relatively low optomechanical coupling of $\gzero/2\pi=220$~kHz~\cite{Safavi-Naeini2013}.  In this work we realize a Si quasi-2D OMC with over $50$-fold improvement in optomechanical back-action per photon, and a much higher thermal conductance ($\times 42$) compared to 1D structures at millikelvin temperatures.  Most importantly, we demonstrate a $Q$-factor of $1.2 \times 10^9$ for the $10$~GHz optomechanically-coupled acoustic mode of the cavity and a $\Ceff$ greater than unity under continuous-wave optical pumping, suitable for realizing applications such as signal transduction of itinerant quantum signals~\cite{Bochmann2013,Andrews2014,balram2016coherent,higginbotham2018harnessing}.


\vspace{2mm}
\noindent\textbf{Results}\\ 
\noindent\textbf{Design of the quasi-2D OMC cavity.} The quasi-2D OMC cavity in this work is designed around the silicon-on-insulator (SOI) materials platform, which naturally provides for a thin Si device layer of a few hundred nanometers in which both microwave-frequency acoustic modes and near-infrared optical modes can be guided in the vertical direction~\cite{Safavi-Naeini2010b}.  Patterning of the Si slab through plasma etching is used to form a nanoscale lattice supporting Bloch waves for both optical and acoustic modes.  We focus on the fundamental guided optical modes of even vector parity about the center of the Si slab ($\sigma_{z} = + 1$).  This choice is motivated by the fact that for a connected lattice of low air-filling fraction the fundamental $\sigma_{z} = + 1$ optical modes are the most strongly guided in the Si slab, greatly reducing their sensitivity to scattering loss.  It is common to refer to these modes as transverse-electric-like (TE-like), as their electric field polarization lies predominantly in the plane of the slab.  For a symmetric Si slab only the acoustic modes of $\sigma_{z} = + 1$ are coupled via radiation pressure to the optical modes of the slab.  

The OMC cavity design consists of three major steps. First, we start with a periodically patterned quasi-2D slab structure with both phononic and photonic bandgaps in which to host the optomechanical cavity. Here we use the `snowflake' crystal with a hexagonal lattice \cite{Safavi-Naeini2010b} as shown in Fig.~\ref{fig1}a. 
The snowflake crystal provides a pseudo-bandgap for TE-like optical guided waves and a full bandgap for all acoustic mode polarizations. Finite-element-method (FEM) simulations of the optical and acoustic modes of the snowflake crystal were performed using the COMSOL software package~\cite{COMSOL}, with nominal snowflake parameters $(a,r,w) = (500,205,75)$~nm and Si slab thickness $t=220$~nm, resulting in a TE-like guided mode photonic band gap extending over optical frequencies of $180$ - $240$~THz (vacuum wavelength $1250$ - $1667$~nm) and an acoustic bandgap covering $8.85$ - $11.05$~GHz. 

Second, we create an in-plane waveguide in the snowflake lattice.  This is done by replacing one row of snowflake unit cells with a customized unit cell.  Waveguiding to this line-defect occurs for photon and phonon modes that lie within the corresponding bandgaps of the surrounding snowflake lattice.  Here, we chose to form the line-defect by replacing one row of snowflakes with a set of `C'-shaped holes. This design is inspired by the one-dimensional nanobeam OMCs reported in Ref.~\cite{Chan2012}. Optomechanical coupling in this sort of design is a result of both bulk (photoelastic)~\cite{Biegelsen1974photoelastic} and surface (moving boundary)~\cite{Johnson2001} effects. 
The `C'-shape allows for large overlap of the acoustic mode stress field with the optical mode intensity in the bulk of the Si device layer, while also focusing the optical mode at the air-Si boundary to increase the moving boundary contribution to the optomechanical coupling. The width of the line-defect, and exact shape and dimension of the `C'-shape were optimized considering several factors: (i) large guided-mode vacuum coupling rate $g_\Delta$~\cite{Safavi-Naeini2010b}, (ii) avoidance of leaky optical resonances of the slab, and (iii) creation of guided acoustic bands with dispersion.  Leaky optical resonances are resonant with the Si slab yet lie above the light cone; imperfections in the fabricated structure can result in large coupling between the guided optical mode of interest and leaky resonances at the same frequency, resulting in large scattering loss.  Acoustic bands of limited dispersion (flat bands) are also susceptible to fabrication imperfections as these acoustic modes tends to localize around small defects resulting in poor overlap with the more extended optical modes (this was a primary difficulty in prior 2D snowflake OMC work~\cite{Safavi-Naeini2013b}).  Photonic and phononic bandstructure diagrams of the optimized waveguide unit cell are shown in Figs.~\ref{fig1}b and \ref{fig1}c, respectively. 
Shaded in light blue are the optical guided-mode bandgap extending from $190$~THz to $210$~THz (vacuum wavelength $1430$ to $1580$~nm) and the acoustic guided-mode bandgap extending from $10$~GHz to $10.6$~GHz.  Abutting these bandgaps and plotted as solid blue curves are the optical and acoustic waveguide bands of interest. 

The final step in the cavity design involves introducing a tapering of the line-defect waveguide properties along the waveguide propagation direction ($x$-axis). Here we utilize a modulation of the `C'-shape parameters that increases quadratically in amplitude with distance along the $x$-axis of the line-defect waveguide from a designated center position of the cavity.  This introduces an approximate quadratic shift of the frequency of the waveguide modes with distance from the cavity center.  For waveguide modes near a band-edge this results in localization of the modes as they are pushed into a bandgap away from the cavity center.  As detailed in App.~\ref{Note1}, a Nelder-Mead simplex search algorithm was used to obtain a tapered cavity structure with simultaneously high optical $Q$-factor and large optomechanical coupling between co-localized optical and acoustic cavity modes. Figures~\ref{fig1}d and \ref{fig1}e display the resulting simulated field profiles of the fundamental optical resonance ($\omegac/2\pi = 194$~THz, $\lambdac = 1550$~nm) and coupled acoustic resonance ($\omegam/2\pi = 10.27$~GHz) of the optimized 2D OMC cavity, respectively.  The co-localized modes have a theoretical vacuum optomechanical coupling rate of $\gzero/2\pi = 1.4$~MHz, and the optical mode has a theoretical scattering-limited quality factor of $\Qoscat = 2.1 \times 10^7$.   



\begin{figure*}[tp]
\begin{center}
\includegraphics[width=2\columnwidth]{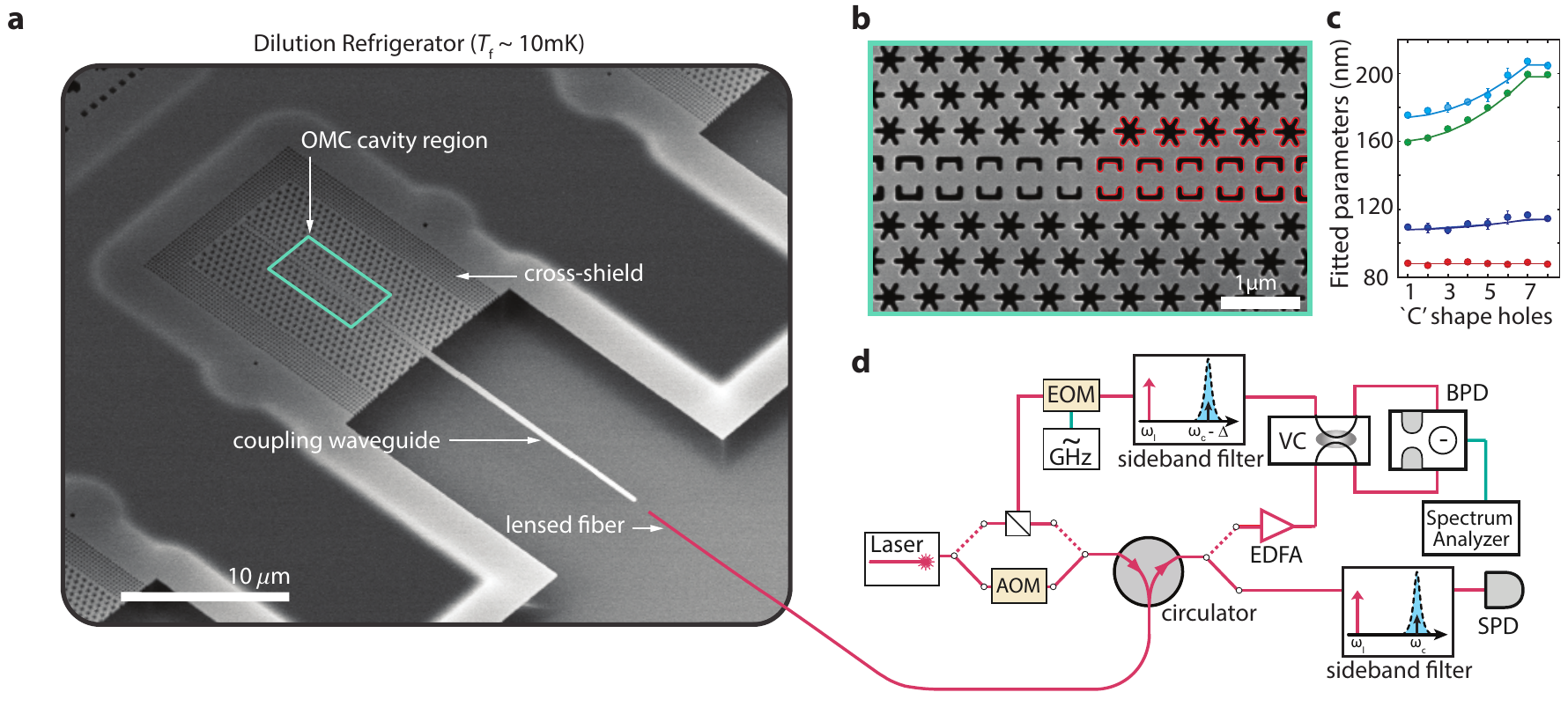}
\caption{\textbf{Device fabrication and measurement setup.} 
\textbf{a}, SEM image of a full quasi-2D snowflake OMC device fabricated on SOI.  This device is an `8-shield device' in which an additional eight periods of cross-structure phononic bandgap shielding is applied at the periphery (see App.~\ref{Note4}). A lensed optical fiber is used to couple light into a tapered on-chip waveguide which is butt-coupling to the quasi-2D snowflake OMC cavity.
\textbf{b}, SEM image of the center of the center of the cavity region. Fitting to the geometries for `C' shape holes and snowflake holes in the cavity region are shown as red solid lines.
\textbf{c}, Measured `C' parameters fit from SEM images of a fabricated cavity (dots) along with their design values (solid curves).  Shown over one half of the cavity (first 8 `C' shapes on the right side of the cavity) are the parameters: $h_\text{i}$ (blue), $h_\text{o}$ (cyan), $w_\text{i}/2$ (red), and $w_\text{o}/2$ (green).
\textbf{d}, Experimental setup for characterization of the quasi-2D snowflake OMC cavity. Multiple optical switches are used to switch between continuous-wave and pulsed optical excitation, and between heterodyne spectroscopy and single photon detection.  Acronyms: Acousto-Optic Modulator (AOM), Electro-Optic Modulator (EOM), Erbium-Doped Fiber Amplifier (EDFA), Variable Coupler (VC), Balanced Photodetector (BPD), Single Photon Detector (SPD). A more detailed schematic and description of the measurement setup is provided in App.~\ref{Note2}.
} 
\label{fig2}
\end{center}
\end{figure*}

Test devices based on this new design were fabricated from a SOI microchip with a $220$~nm thick Si device layer and an underlying $3$~$\mu$m buried oxide layer. A scanning electron microscope (SEM) image of a fabricated 2D OMC cavity and optical waveguide for coupling light into the structure are shown in Figs.~\ref{fig2}a and ~\ref{fig2}b. Several iterations of fabrication were performed in order to improve the fidelity of the fabrication with respect to the design structure.  Between fabrication iterations SEM images of devices were analyzed to determine the fabricated geometrical parameters of the cavity structure; this information was fed back into the next fabrication iteration in order to realize devices with a geometry as close as possible to the simulation-optimized design parameters. An example of fitted `C'-shape and snowflake holes are shown as red solid lines in the SEM image of Fig.~\ref{fig2}b, with corresponding fitted cavity parameters plotted in Fig.~\ref{fig2}c.
\par

\begin{figure}[tp]
\begin{center}
\includegraphics[width=\columnwidth]{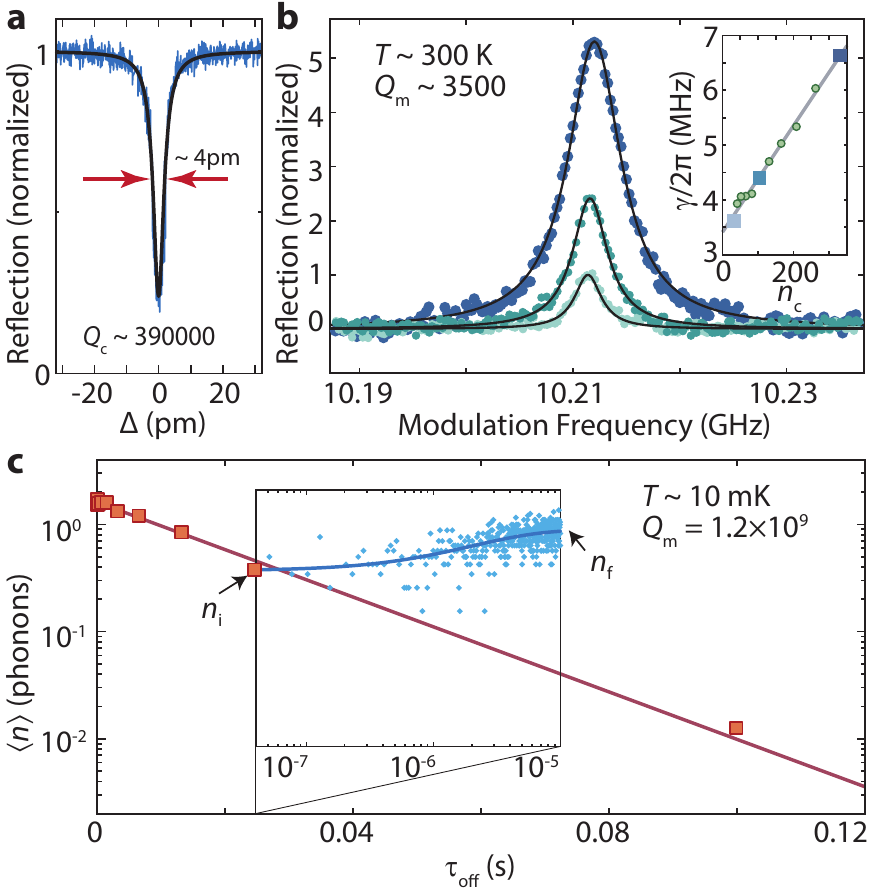}
\caption{\textbf{Optomechanical characterization.} 
\textbf{a}, Optical spectrum of a quasi-2D snowflake OMC cavity measured using a swept laser scan.
\textbf{b}, Pump-probe measurement of the mechanical mode spectrum of interest centered around $\omegam/2\pi = 10.21$~GHz for different optical pump powers (from lowest to highest peak reflection: $\ncav = 33$, $\ncav = 104$, and $\ncav = 330$). Inset: measured mechanical mode linewidth versus $\ncav$.  The device measured in \textbf{a} and \textbf{b} is a cavity with zero additional acoustic shield periods (zero-shield device). 
\textbf{c}, Pulsed ringdown measurement of a quasi-2D snowflake OMC cavity with 8 periods of additional cross-structure shielding (8-shield device).  The measured optomechanical parameters of this device are $(\kappa, \kappae, \gzero, \omegam, \gammanotO) = 2\pi(1.187$~GHz, $181$~MHz, $1.182$~MHz, $10.02$~GHz, $8.28$~Hz$)$. The phonon occupancy at the beginning of each $10$~$\mu$s optical pulse, $n_\text{i}$ (orange squares), is measured using a peak power corresponding to $\ncav = 60$. Inset: mode heating within the optical pulse, with blue circles corresponding to data and the solid line a fit to the heating curve. $n_{\text{f}}$ is the mode occupancy at the end of the optical pulse, used here to check for consistent excitation across the different inter-pulse $\Toff$ measurements. 
} 
\label{fig3}
\end{center}
\end{figure}

\vskip 0.2in
\noindent\textbf{Optomechanical coupling and mechanical damping.} Fabricated devices were characterized both at room temperature ($300$~K) and at cryogenic temperatures inside a fridge ($\Tf=10$~mK). A simplified schematic of the optical measurement set-up is shown in Fig.~\ref{fig2}d.  Room temperature testing was performed using a dimpled optical fiber taper to evanescently couple light into and out of a chip-based Si coupling waveguide~\cite{Groeblacher2013a}; each Si coupling waveguide is butt-coupled to a corresponding OMC cavity as shown in Fig.~\ref{fig2}a (also see App.~\ref{Note3}).  A typical optical spectrum from one of the quasi-2D OMC cavities is displayed in Fig.~\ref{fig3}a, showing a fundamental optical resonance at a wavelength of $\lambdac = 1558.8$~nm with a loaded (intrinsic) optical $Q$-factor of $\Qo=3.9 \times 10^5$ ($\Qoi=5.3 \times 10^5$).  

In order to measure the coupled acoustic resonance(s) of the OMC cavity we used a pump-probe scheme involving a laser pump tone of frequency $\omegap$ fixed near the red motional sideband of the optical cavity ($\Delta \equiv \omegac-\omegap \approx +\omegam$).  In this scheme, electro-optic modulation of the pump laser is used to create a weak laser probe tone of frequency $\omegas$ that we tune across the OMC cavity resonance by sweeping the modulation frequency~\cite{Safavi-Naeini2011}.  The reflected pump and probe laser tones from the OMC cavity contain the driven response of those optomechanically-coupled acoustic modes that are in two-photon resonance with the laser drive fields (i.e., $\omegam = \omegas-\omegap$).  Coherent detection of the beating of the pump and probe laser tones on a high speed photodetector produces a spectrum of the coupled acoustic modes.  For the new quasi-2D OMC cavity design we found a single, dominantly-coupled acoustic mode around $\omegam/2\pi \approx 10.2$~GHz.  A plot of the measured acoustic mode spectrum at several optical pump powers is shown in Fig.~\ref{fig3}b for the device of Fig.~\ref{fig2}a. Optomechanical back-action from the pump laser can be seen to broaden the acoustic resonance; a plot of the fit resonance linewidth ($\gammatot$) versus intra-cavity photon number of the pump laser tone ($\ncav$) is shown as an inset to Fig.~\ref{fig3}b.  From the slope of the back-action-broadened linewidth versus $\ncav$ we extract a vacuum coupling rate of $\gzero/2\pi = 1.09$~MHz, close to the simulated optimum value of $1.4$~MHz.

Following initial room temperature measurements, the new optimized quasi-2D OMC cavities were tested at millikelvin temperatures.  We measured the intrinsic mechanical damping rate, backaction cooling, and heating dynamics of the OMC cavities in a dilution refrigerator (DR) with a temperature of $T_\text{f} \approx 10$~mK at the base plate connected to the mixing chamber. The $10$~mm by $5$~mm sample containing an array of devices was directly mounted on a copper mount attached to the mixing chamber plate, and a 3-axis stage was used to align a lensed optical fiber to the tapered on-chip coupling waveguide of a given device under test (see Fig.~\ref{fig2}a and App.~\ref{Note3}).

We measured the intrinsic mechanical $Q$-factor of the quasi-2D OMC devices at millikelvin temperatures using a pulsed optical scheme in which $10$-microsecond-long optical pulses excite and read-out the energy in the acoustic mode (for details see the Methods section).  By varying the delay between the optical pulses, this technique allows for the evaluation of the acoustic energy ringdown while the laser light field is off~\cite{Meenehan2015b,MacCabe2019}.  In Fig.~\ref{fig3}c we show the measured ringdown curve for a $10$~GHz acoustic mode of a quasi-2D snowflake OMC cavity with an additional acoustic shield.  In order to increase the acoustic isolation of the acoustic cavity mode, on this device we added a periphery consisting of 8 periods of a cross-structure phononic bandgap shield (see App.~\ref{Note4}).  Fitting of the initial mode occupancy at the beginning of an optical pulse ($n_\text{i}$) versus inter-pulse delay ($\Toff$) yields an intrinsic acoustic energy decay rate of $\gammanotO/2\pi=8.28$~Hz, corresponding to a mechanical $Q$-factor of $\Qm = 1.2 \times 10^{9}$.  This large mechanical $Q$-factor is consistent with other measurements of Si nanobeam OMC cavities at millikelvin temperatures~\cite{MacCabe2019}, and is thought to result from a suppression of acoustic absorption from near-resonant two-level system defects~\cite{Phillips1987}.   

\begin{figure}[tp] 
\begin{center}
\includegraphics[width=\columnwidth]{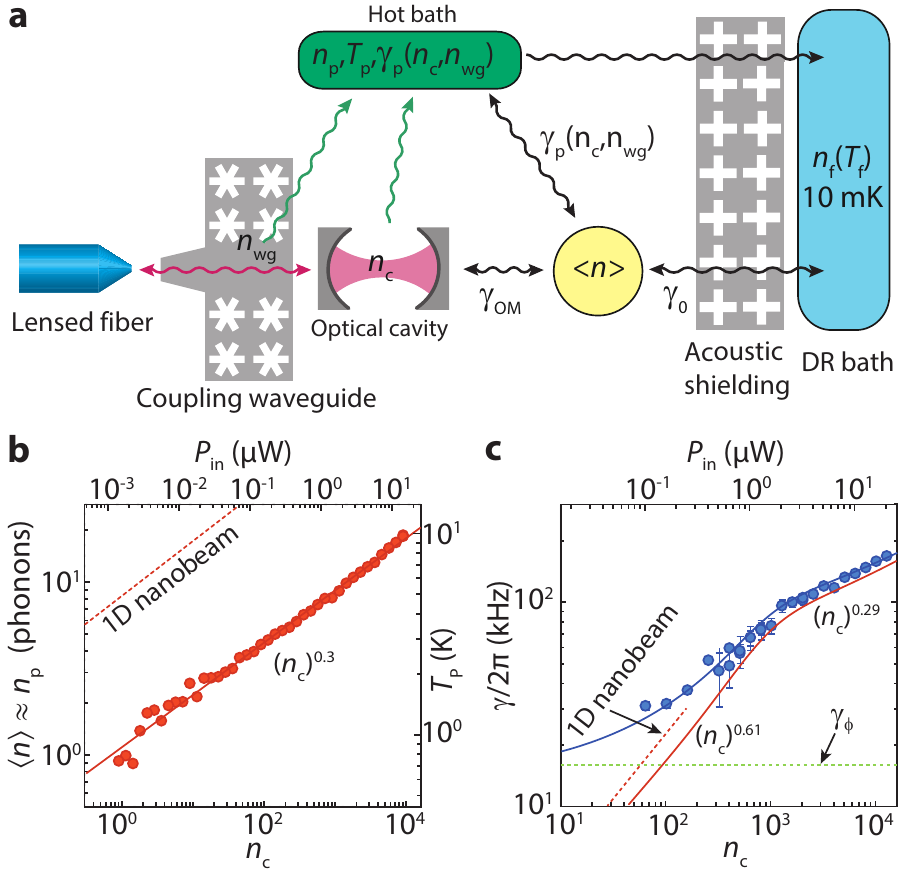}
\caption{\textbf{Optical-absorption-induced hot bath.} \textbf{a}, Diagram illustrating the proposed model of heating of the mechanics due to optical absorption and the various baths coupled to the localized mechanical mode.
\textbf{b}, Plot of measured $\nbathp$ versus $\ncav$. The solid line is a fit to the data giving $\nbathp = (1.1) \times \ncav^{0.3}$. 
Dashed line indicates $\nbathp$ versus $\ncav$ for a 1D nanobeam device measured in Ref.~\cite{MacCabe2019} for comparison. 
\textbf{c}, Plot of measured linewidth $\gammatot$ (blue dots) versus $\ncav$. The blue solid line is a power-law fit to the data, where in the low $\ncav$ regime $\gammatot/2\pi \approx \gammaphi/2\pi + \gammap/2\pi = 14.54~\text{kHz} + (1.1~\text{kHz}) \times \ncav^{0.61}$ and in the high $\ncav$ regime
$\gamma/2\pi = 23.91~\text{kHz} + (9.01~\text{kHz}) \times \ncav^{0.29}$. The red solid curve is the resulting fit for $\gammap$ by itself.  For comparison, the dashed red curve is a plot of $\gammap$ versus $\ncav$ for a 1D nanobeam device from Ref.~\cite{MacCabe2019}.  In \textbf{b} and \textbf{c} the quasi-2D OMC cavity measurements are for the same 8-shield device as in Fig.~\ref{fig3}c.
}
\label{fig4}
\end{center}
\end{figure}

\vskip 0.2in
\noindent\textbf{Optical-absorption-induced hot bath.} While the acoustic mode $Q$-factor measured in ringdown measurements is promising for certain quantum memory applications~\cite{pechal2018superconducting,hann2019hardware}, it is measured with the laser pump off. The prospects for performing coherent quantum operations between photons and phonons depends critically on the ability to minimize unwanted heating and damping of the acoustic mode due to parasitic effects resulting from optical absorption in the presence of an applied laser field. A model for the heating and damping in Si OMC cavities, first proposed in Ref.~\cite{Meenehan2014}, is illustrated in Fig.~\ref{fig4}a.  In this model, the acoustic mode of the OMC cavity is weakly coupled to the surrounding DR environment at a rate $\gammanotO$, while simultaneously being coupled to an optically-generated hot bath at a rate $\gammap$.  The source of the hot bath in Si OMC devices is thought to be due to linear optical absorption via electronic defect states at the surface of the etched Si~\cite{Stesmans1996,Borselli2007}.  A model that treats the hot bath as consisting of high frequency phonons lying above the acoustic bandgap of the OMC structure predicts -- assuming a phonon density of states corresponding to that of a 2D plate and weak coupling to the localized acoustic cavity mode via 3-phonon scattering -- acoustic mode heating that scales as $\ncav^{1/3}$ and damping that scales as $\ncav^{2/3}$~\cite{MacCabe2019}.



Here we explore the optically-induced parasitic heating and damping for the quasi-2D OMC cavity.  As for the previous 1D nanobeam measurements, optical measurements were performed using a continuous-wave pump laser tuned to the optical cavity resonance ($\Delta = 0$) to avoid optomechanical back-action cooling and damping (see Methods). Results of the optically-induced heating and damping of the $10$~GHz breathing-like mode of the 2D OMC cavity are displayed in Fig.~\ref{fig4}b and ~\ref{fig4}c, respectively.  In Fig.~\ref{fig4}b the inferred hot bath occupancy $\nbathp$ (left vertical axis) and the corresponding bath temperature $\Tbathp$ (right vertical axis) are plotted versus intra-cavity photon number $\ncav$ (lower horizontal axis) and the corresponding input power $\Pin$ (upper horizontal axis).  We find that the hot bath occupancy is accurately fit by the power-law, $\nbathp = (1.1)\times \ncav^{0.3}$.  This is the same power-law scaling as found for 1D nanobeam OMCs in Ref.~\cite{MacCabe2019} (red dashed line in Fig.~\ref{fig3}b); however, the overall magnitude of the hot bath occupancy has substantially dropped for the quasi-2D cavity by a factor of 7.2.  Using a modified thermal conductance model that assumes ballistic phonon transport and a power-law dependence on temperature consistent with our measurements, $\thermcond = \epsilon (\Tbathp)^{\alpha=2.3}$, FEM numerical simulations of the 1D and 2D cavity structures (see App.~\ref{Note5}) predict a greatly enhanced thermal conductance coefficient for the quasi-2D cavity ($\epsilontwoD/\epsilononeD = 42$).  This yields a hot bath occupancy ratio between nanobeam and quasi-2D cavities of 6.2, in good correspondence to the measured value.  In alignment with our design strategy, the reduction in mode heating according to this model is primarily the result of a geometric effect due to the fact that the quasi-2D cavity is connected to the surrounding chip structure over a larger in-plane solid angle than that of the 1D nanobeam, and thus has a larger number of phonon modes to carry heat away (i.e., a larger thermal conductance).

In Fig.\ref{fig4}c we plot the spectral linewidth of the breathing mode of the quasi-2D cavity, determined in this case by measuring the acoustic mode thermal noise spectrum imprinted on the reflected laser pump field using a balanced heterodyne receiver (see Fig.~\ref{fig2}d and Methods).  In such a measurement, with resonant pump laser field ($\Delta = 0$), the acoustic mode linewidth will be the sum of the intrinsic energy decay rate $\gammanotO$, the optical absorption bath-induced damping $\gammap$, and any pure dephasing effects (frequency jitter) of the acoustic resonance $\gammaphi$. Referring to Fig.\ref{fig4}c, we see that the linewidth dependence on $\ncav$ can be broken into three different regimes: (i) at the lowest powers ($\ncav \lesssim 10$) the linewidth saturates to a constant value given by $\gammaphi$ ($\gammanotO$ is entirely negligible on this scale), (ii) a low-power regime ($100 < \ncav < 1000$) with a relatively strong dependence of linewidth on optical power, and (iii) a high-power regime ($\ncav > 1000$) with a second, weaker dependence of linewidth on optical power.  Fitting the low-power regime with a power-law dependence on $\ncav$, we find $\gammatot/2\pi = \gammaphi/2\pi + (1.1$~kHz$)\times \ncav^{0.61}$, with $\gammaphi/2\pi = 14.54$~kHz.  At these lower powers we find a power-law scaling and overall magnitude of damping of the breathing mode of the quasi-2D OMC cavity which is close to that for the 1D nanobeam cavities of Ref.~\cite{MacCabe2019} (see the dashed red curve in Fig.~\ref{fig4}c).  In the high-power regime we find a fit given by $\gammatot/2\pi = 23.91$~kHz$ + (9.01$~kHz$)\times \ncav^{0.29}$, with a power-law exponent that is approximately half that in the low-power regime.  Determining the exact mechanism of the $\gammap$ slow down versus $\ncav$ (or indirectly $\Tbathp$) in the high-power regime is outside the scope of this Article, however, possibilities include a change in the phonon scattering rate with increasing phonon frequency in the nanostructured Si film~\cite{chen2000particularities,cross2001elastic} or a transition from Landau-Rumer scattering to Akhiezer-type damping as the effective bath temperature rises~\cite{akhiezer1939sound}.   

Before moving on to measurements of back-action cooling in the quasi-2D OMC cavity, we note one important distinction between the geometry of the optical coupling in the new 2D devices in comparison to previously studied 1D nanobeam devices. Whereas in the 1D nanobeam devices the coupling waveguide is evanescently coupled to the OMC cavity -- and is thus not in direct mechanical contact with the nanobeam cavity -- in the quasi-2D devices the optical coupling waveguide is physically connected to one end of the OMC cavity region (see Fig.~\ref{fig2}a).  Optical absorption in the coupling waveguide of the quasi-2D OMC devices may thus also lead to heating of the acoustic cavity mode.  This effect is further corroborated by FEM simulations, detailed in App.~\ref{Note6}, that show that a weak cavity is formed between the end of the coupling waveguide and the quasi-2D OMC cavity.  Optical absorption of the input power ($\Pin$) in the coupling waveguide can be modelled as an effective waveguide photon number, $\nwg$, where $\nwg=\beta\Pin$ for some fixed constant $\beta$ independent of cavity detuning. Assuming that the dependence of $\nbathp$ on $\nwg$ is the same as that for $\ncav$, we can write $\nbathp(\ncav,\Pin) = \nbathp(\ncav + \nwg)$.  Similarly, $\gammap(\ncav,\Pin) = \gammap(\ncav + \nwg)$.  In the measurements above with resonant pumping at $\Delta=0$, very small input powers were required to build up large intra-cavity photon numbers, and as such $\ncav \gg \nwg$.  In what follows, where we perform back-action cooling with $\Delta = \omegam \gg \kappa$, the input power required to yield a given $\ncav$ is much larger and $\nwg$ cannot be ignored.




\vskip 0.2in
\noindent\textbf{Effective quantum cooperativity.} The ability of a cavity optomechanical systems to perform coherent quantum operations between the optical and mechanical degrees-of-freedom requires both large cooperativity $C \equiv \gammaOM/\gammab$ and a mechanical mode thermal occupancy $\nbar < 1$ (the thermal noise in the optical mode is assumed negligible)~\cite{Hill2012,Andrews2014}. Here $\gammab$ represents the total coupling rate of the mechanical system to its various thermal baths, which in the case of the Si OMC cavities is given by $\gammab = \gammanotO+\gammap$.  The relevant figure-of-merit is then the effective quantum cooperativity $\Ceff \equiv C/\nbath$~\cite{Aspelmeyer2014}, where $\nbath$ is the total effective bath occupancy defined by $\gammab \nbath \equiv \gammanotO (\nnaught+1) + \gammap (\nbathp+1)$.  In this relation for $\gammab\nbath$ the `+1' terms correspond to spontaneous decay and $\nnaught \lesssim 10^{-3}$ is the bath occupancy in the surrounding chip region of the OMC cavity (see App.~\ref{Note7}).  In general, $\gammab\nbath$ would also include any dephasing $\gammaphi$.  As $\gammanotO \ll \gammap$ for the optical powers used in this work and $\nnaught \ll 1$, in what follows $\gammab \nbath \approx \gammap (\nbathp+1)$.   


\begin{figure}[tp] 
\begin{center}
\includegraphics[width=\columnwidth]{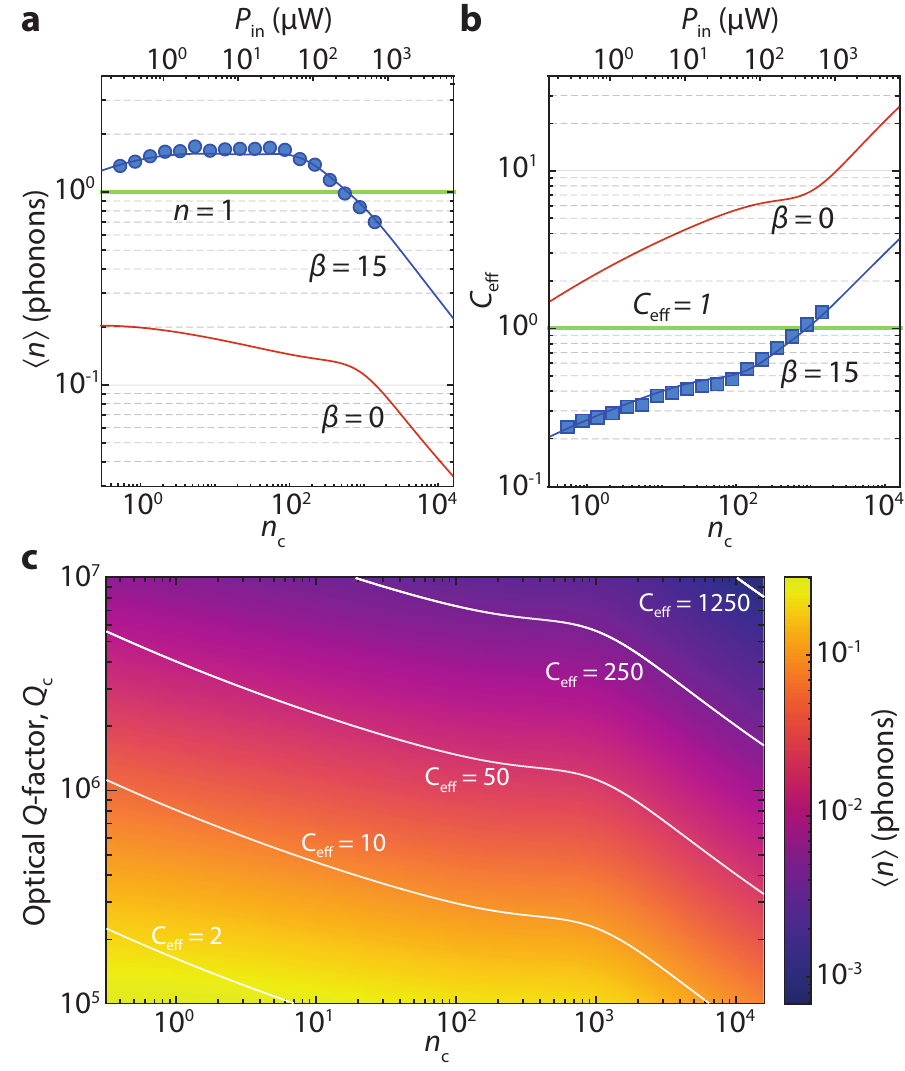} 
\caption{\textbf{Phonon occupancy and effective quantum cooperativity.} \textbf{a}, Plot of the measured occupancy $\nbar$ in the acoustic mode at $\omegam/2\pi = 10.02$~GHz versus laser pump power in units of intra-cavity photons, $\ncav$, and optical power coupled into coupling waveguide, $\Pin$. Filled blue circles are measured data.
\textbf{b}, Plot of quantum cooperativity $\Ceff$ versus $\ncav$ and $\Pin$.  The $\Ceff$ data points (filled blue squares) are inferred using Eq.~(\ref{eq:nbarcool}) from the measured $\nbar$ in panel \textbf{a} along with the fit to the measured $\nbathp$ and $\gammap$ from Fig.~\ref{fig4}.  In both \textbf{a} and \textbf{b} the solid curves are theoretical plots calculated using the power-law fits of $\gammap$ and $\nbathp$ considering optical heating from $\ncav$ alone ($\beta=0$, red solid line curve) and including direct heating from $\Pin$ with $\beta = 15$~$\mu$W$^{-1}$ (blue solid curve).
\textbf{c}, Estimated back-action cooled phonon occupancy $\nbar$ for a quasi-2D OMC device with the same properties as measured in panel \textbf{a} save for a modified $Q$-factor and no waveguide heating ($\beta=0$). White solid curves delineate contours of constant quantum cooperativity $\Ceff$. In panels \textbf{a} and \textbf{b}, quasi-2D OMC cavity measurements are for the same device as in Fig.~\ref{fig3}c and Fig.~\ref{fig4}.
}
\label{fig5}
\end{center}
\end{figure}

A measurement of the quantum cooperativity $\Ceff$ can be made by observing the cooled mechanical occupancy under optical back-action cooling of the coupled mechanical mode,

\begin{equation}
\nbar = \frac{\gammap \nbathp + \gammanotO \nnaught}
{\gammanotO + \gammaOM + \gammap} = \frac{\nbath - 1}{C + 1} \overset{\nbath,C \gg 1}{\approx} \frac{1}{\Ceff},
\label{eq:nbarcool}
\end{equation}

\noindent where we have implicitly assumed that the optical pump laser responsible for producing back-action damping $\gammaOM$ has a zero effective noise occupancy.  Back-action cooling is most efficient in the resolved sideband limit ($\kappa/2\omegam < 1$) when the optical pump is applied on the red-detuned motional sideband of the cavity, $\Delta = \omegam$.  In Fig.~\ref{fig5}a we show the measured cooling curve of the quasi-2D snowflake cavity under continuous-wave optical pumping at $\Delta=\omegam$.  We infer the measured $10.2$~GHz acoustic cavity mode occupancy (blue dots) from calibration of the photon counts of the anti-Stokes sideband of the reflected optical pump laser (see Fig.~\ref{fig2}d and Methods).  



We can also predict the back-action cooling curve of the acoustic mode based upon our independent measurements of $\gammaOM$ (Fig.~\ref{fig3}b), $\gammap$ (Fig.~\ref{fig4}b) and $\nbathp$ (Fig.~\ref{fig4}c) versus $\ncav$.  Using Eq.~(\ref{eq:nbarcool}) we plot the theoretical acoustic mode cooling versus $\ncav$ (i.e., $\beta=0$) as a solid red curve in Fig.~\ref{fig5}a.  Not only does the $\beta=0$ theoretical curve predict substantially more cooling of the acoustic mode than measured, but the shape of the measured and theoretical curves are also quite different.  As alluded to at the end of the previous section, one significant difference between the back-action cooling measurements and the measurements of $\nbathp$ and $\gammap$ is that the cooling measurements are performed at $\Delta=\omegam$, requiring a hundred-fold increase in the optical pump power to reach a given intra-cavity photon number $\ncav$.  For reference we have plotted the corresponding input power $\Pin$ on the top horizontal axes of Figs.~\ref{fig4}b, \ref{fig4}c and \ref{fig5}a.  Adding an additional waveguide photon number $\nwg=\beta\Pin$ to the intra-cavity photon number $\ncav$ in determining $\nbathp$ and $\gammap$, we find a modified cooling curve ($\beta=15$~$\mu$W$^{-1}$; solid blue curve) that fits both the magnitude and shape of the measured cooling curve.  In Fig.~\ref{fig5}b we plot the corresponding quantum cooperativity curve showing that $\Ceff$ reaches a value as high as $1.3$, with a continuing trend well above unity at even larger input powers (in our experiments we were limited to $\ncav \lesssim 1000$ for back-action cooling due to large detuning and limited laser power).               

Although we have achieved $\Ceff \approx 1.3 > 1$ under continuous-wave optical driving in the newly designed quasi-2D OMC cavities, looking forward, significant further increases can be achieved. One clear method is to thermally decouple the input coupling waveguide from the OMC cavity in order to eliminate the parasitic heating from $\nwg$.  This can be accomplished, for instance, by using evanescent side-coupling instead of butt-coupling of the coupling waveguide to the cavity.  A second approach to dramatically improving $\Ceff$ is through improvements in optical quality factor, where similar quasi-2D planar photonic crystal devices have already been demonstrated with optical $Q$-factors approaching $10^7$~\cite{Sekoguchi14}. 
In Fig.~\ref{fig5}c we estimate achievable $\nbar$ and $\Ceff$ as a function of  $\ncav$ and $\Qo$ assuming that $\nwg$ has been successfully eliminated.  We find that for a $Q$-factor of $\Qo = 3.90 \times 10^5$, equal to that of the zero-shield device of Fig.~\ref{fig3}a, it should be possible to reach $\nbar\approx 0.1$ and $\Ceff \approx 5$ for optical pump powers at the single photon level. 


\vspace{2mm}
\noindent\textbf{Discussion}\\ 
\noindent In conclusion, we have presented the design, fabrication and characterization of a new quasi-2D optomechanical crystal cavity with a breathing-like acoustic mode of $10$~GHz frequency and large vacuum optomechanical coupling rate $\gzero/2\pi \gtrsim 1$~MHz. By employing an engineered 2D phononic and photonic bandgap material in which to host the OMC cavity, the acoustic breathing-like mode of interest is well protected from its environment, while phonon modes above the acoustic bandgap serve as additional channels for removing heat from the cavity region.  Through this dual role of the 2D bandgap structure, we demonstrate at millikelvin temperatures a localized acoustic cavity mode with intrinsic $Q$-factor of $1.2 \times 10^{9}$ and a greatly increased ($\times 68$) thermal conductance between the cavity and the cold bath reservoir of the surrounding chip compared to previous 1D nanobeam OMC devices.  These properties of the quasi-2D OMC cavity allow us to achieve a quantum cooperativity $\Ceff > 1$ under continuous-wave optical pumping.  They also point the way for significant further improvements in quantum cooperativity through modification in the optical coupling geometry.  

This result ushers forth a variety of quantum optomechanical applications using chip-scale optomechanical crystals.  In particular, in the case of hybrid superconducting and acoustic microwave quantum circuits~\cite{Arrangoiz_Arriola2016,arrangoiz2018coupling,Kalaee2019,Arrangoiz_Arriola2019}, optomechanical devices that can operate in continuous mode in the high $\Ceff$ regime at millikelvin temperatures would enable bi-directional conversion of itinerant optical and microwave quantum signals, forming the critical interface necessary to realize an optical quantum network~\cite{Kimble2008} of superconducting quantum circuit nodes. These advances may also allow quantum optomechanical measurements to be performed at even lower temperatures than currently possible with OMC cavities, which given their demonstrated ultralow rates of intrinsic acoustic dissipation~\cite{MacCabe2019}, will allow for further studies of theories related to gravitationally-induced decoherence~\cite{blencowe2013effective,pikovski2015universal} and nonlinearities and dephasing in mechanical systems~\cite{atalaya2016nonlinear}.

\vspace{2mm}
\noindent\textbf{Methods}\\ 
\noindent\textbf{Device fabrication.} The devices were fabricated from a SOI wafer (SEH, $220$~nm Si device layer, 3~$\mu$m buried-oxide layer) using electron-beam lithography followed by inductively coupled plasma reactive ion etching (ICP/RIE). The Si device layer is then masked by photoresist to define a `trench' region of the chip to be etched and cleared, to which a lens fiber can access the chip coupling waveguides. In the unprotected trench region of the chip, the buried-oxide layer is removed with a highly anisotropic plasma etch, and the handle Si layer is removed to a depth of 120 $\mu$m using an isotropic plasma etch. The devices were then released in vapor-HF and cleaned in a piranha solution (3-to-1 H$_{2}$SO$_{4}$:H$_{2}$O$_{2}$) before a final diluted-HF etch to remove any surface oxides.

\vskip 0.2in
\noindent\textbf{Device characterization.} Fabricated devices are characterized using a fiber-coupled, wavelength-tunable external cavity diode laser.  The laser light is sent through a $50$~MHz-bandwidth tunable fiber Fabry-Perot filter (Micron Optics FFP-TF2) to reject laser phase noise at the mechanical frequency. After this prefiltering, the light path can be switched by 2$\times$2 optical switches between two paths: (i) a `balanced heterodyne spectroscopy' path with a high-speed photodetector for performing balanced heterodyne detection of the acoustic mode, and (ii) a `photon counting path' with a single photon detector for performing either pulsed or continuous-wave phonon counting of the acoustic mode. 

For the balanced heterodyne spectroscopy path, a $90:10$ beam-splitter divides the laser source into local oscillator (LO, 90\%) and signal (10\%) beams. The LO is modulated by an electro-optic modulator (EOM) to generate a sideband at $\delta/2\pi = 50$~MHz from the mechanical frequency.  One of the modulated sidebands of LO is selected using a high-finesse tunable Fabry-Perot filter before recombining it with the signal for detection. The signal path is sent through a variable optical attenuator and then to an optical circulator which directs the signal to the device under test in the dilution refrigerator.  A lensed optical fiber in the dilution refrigerator is used to couple light into and out of the devices. The reflected signal beam is recombined with the LO using a variable optical coupler, the outputs of which are sent to a balanced photodetector (BPD). 

For the photon counting path, the laser is directed via 2$\times$2 mechanical optical switches into a `high-extinction' branch consisting of an acousto-optic modulator (AOM; $20$~ns rise and fall time, $50$~dB on-off ratio) and two Agiltron NS switches (Ag.; $100$~ns rise time, $30$~$\mu$s fall time, total of $36$~dB on-off ratio), which are driven by a digital delay generator to generate high-extinction-ratio optical pulses. The digital delay generator is used to synchronize the switching of the AOM and Agiltron switches with the time-correlated single-photon-counting (TCSPC) module which is connected to a single-photon detector (SPD).  The SPDs used in this work are amorphous WSi-based superconducting nanowire single-photon detectors. The tunable fiber Fabry-Perot filters used for both pre-filtering the pump and filtering the motionally-generated sidebands have a bandwidth of $50$~MHz, a free-spectral range of $20$~GHz, and a tuning voltage of $\leq 18$~V per free-spectral range. Each of the filters provide $40$~dB extinction at $10$~GHz offset from the transmission peak. 

\vskip 0.2in
\noindent\textbf{Calibration of $\ncav$ and $\gzero$.}
The photon number $\ncav$ at a given power and detuning depends on the single pass fiber-to-waveguide coupling efficiency $\eta_{\text{cpl}}$ and cavity extrinsic coupling efficiency $\eta_{\kappa} = \kappae/\kappa$. The fiber-to-waveguide coupling efficiency $\eta_{\text{cpl}}$ was determined by measuring the reflection level far off-resonance from the optical cavity using a calibrated optical power meter, and was found to be $\eta_{\text{cpl}}$ = 0.6 for the eight-shield device of this work. The cavity extrinsic coupling efficiency $\eta_{\kappa}$ was measured by placing the frequency of the pump laser far off-resonance and using a vector network analyzer (VNA) to drive an EOM to sweep an optical sideband through the cavity frequency. The optical response is measured on a high-speed photodiode that is connected to the VNA signal port. The amplitude and phase response of the cavity are obtained and fit to determine $\eta_{\kappa}$ and $\kappa$. With these two parameters we determine $\ncav$ for a given input power,

\begin{equation}
\ncav = \frac{P_\text{in}}{\hbar \omegap}\frac{\kappae}{\Delta^2 + (\kappa/2)^2},
\label{eqn:nc_expression}
\end{equation}
\par

\noindent where $\omegap$ is the pump laser frequency. To extract the vacuum optomechanical coupling rate $\gzero$, we measure the acoustic mode linewidth versus optical power for $\Delta=\omegam$.  The slope of the linewidth versus the calibrated $\ncav$ yields $4 \gzero^2/ \kappa$, from which we determine $\gzero$. At millikelvin temperatures we use the measured per-phonon scattering rate, $\gammaSB$, in a similar fashion to determine $\gzero$ (see below).  

\vskip 0.2in
\noindent\textbf{Ringdown measurements.}
The intrinsic mechanical $Q$-factor of the quasi-2D OMC devices is measured using a ringdown technique.  The ringdown measurement we employ uses pulsed optical excitation and photon counting techniques as presented previously in Refs.~\cite{Meenehan2015b,MacCabe2019}. In the simplest version, the measurement relies on generating a train of optical laser pulses detuned to the lower-frequency motional sideband of the optical cavity ($\Delta=\omegam$), with excitation and read-out performed by the same pulse. Excitation of the acoustic mode into a low-occupancy thermal state (a few phonons) is provided by the optical absorption heating during the pulse.  Read-out of the acoustic mode occupancy is performed by photon counting of the anti-Stokes sideband photons of the pump as described below in the discussion of the back-action cooling measurements.  

For ringdown measurements in this work we use a train of $10$-microsecond-long high-extinction optical laser pulses with the laser frequency tuned to the red-motional-sideband of the OMC cavity resonance ($\Delta = \omegam$). The pulses are generated as described above in the description of the photon counting path of the measurement set-up, with an on-off extinction exceeding $80$~dB and rise and fall times of approximately $20$~ns.  We used a peak pulse power corresponding to $\ncav=60$.  For each $\Toff$ we average over many optical pulses the ratio of the inferred mode occupancy within the first $25$~ns time bin of the optical pulse ($\nmi$) to that of the inferred occupancy in the last $25$~ns time bin of the optical pulse ($\nmf$).  The inter-pulse delay $\Toff$ is varied to map out the decay of the acoustic mode energy in between optical pulses.

\vskip 0.2in
\noindent\textbf{Measurement of hot bath occupancy $\nbathp$.}
Measurements of the occupancy of the hot bath are performed using the photon counting path of the measurement set-up, with the pump laser operated in continuous-wave (i.e., no modulation) and tuned to cavity resonance ($\Delta = 0$). The reflected pump laser from a device under test is filtered at an offset equal to the mechanical mode frequency to separate the motionally-generated anti-Stokes sideband from the pump.  This filtered signal is then sent to the SPD for detection. The measured photon count rate for this pump laser detuning is given by $\Gamma(\Delta=0) = \eta (\kappa/2 \omegam)^2 \gammaOM \nbar$, where $\eta$ is the total optical detection efficiency of sideband photons, $\kappa$ is the total optical cavity mode decay rate and $\gammaOM=4\gzero^2\ncav/\kappa$ is the parametric optomechanical coupling rate. The observed count rate -- in conjunction with calibration of $\eta$ and independent measurement of $\kappa$, $\omegam$, and $\gammaOM$ -- is then used to extract the mode occupancy of the coupled breathing-like mode, $\nbar$.  Due to the lack of optomechanical back-action damping for $\Delta=0$ laser detuning, this measured occupancy is a close approximation to the hot bath occupancy $\nbathp$ at power levels where the effective hot bath temperature $\Tbathp$ is much greater than the base temperature to which the mode thermalizes with no applied laser, $\Tbathnaught \approx 63$~mK (see App.~\ref{Note7}).  

\vskip 0.2in
\noindent\textbf{Back-action cooling measurements.}
Back-action cooling measurements are also performed using the photon counting path of the measurement set-up, but with the optical pump laser detuned to the red motional sideband of the optical cavity resonance ($\Delta = \omegam$).  An accurate measure of the acoustic phonon mode occupancy for the back-action cooling measurements can be determined by using vacuum noise as a reference meter. The measured photon count rate for a red- or blue-detuned pump laser is given by~\cite{Meenehan2015b}:

\begin{align}
\Gamma(\Delta = \pm \omegam) = \Gamma_{\text{DCR}} + \Gamma_{\text{pump}} + \gammaSB(\langle n \rangle + \frac{1}{2}(1 \mp 1)),
\end{align}

\noindent where $\Gamma_{\text{DCR}}$ is the dark count rate of the SPD, $\Gamma_{\text{pump}}$ is the count rate due to any bleed-through from the sideband filtering of the pump laser, and $\gammaSB = \eta_{\text{det}} \eta_{\text{cpl}} \eta_{\kappa} \gammaOM$ is the detected photon scattering rate per phonon on the SPD. Here $\eta_{\text{det}}$ is the measured overall detection efficiency of the set-up, including insertion losses in the fibers inside and outside of the dilution refrigerator, fiber unions, fiber circulator, and the detection efficiency of the SPD ($\etaSPD$). We then calibrate $\gammaSB$ using a pulsed blue-detuned laser pump ($\Delta = - \omegam$). The inter-pulse delay ($\Toff$) of the blue-detuned pulses was selected to be much longer than the intrinsic damping time of the acoustic mode such that at the beginning of the pulse $\nbar \approx \nnaught \ll 1$ and the sideband photon count rate is set by spontaneous vacuum scattering of the pump, $\Gamma \approx \gammaSB$ ($\Gamma_{\text{DCR}} + \Gamma_{\text{pump}}$ was confirmed to be much smaller than $\gammaSB$). Normalizing the measured count rate by $\gammaSB$ in the back-action cooling measurements yields the calibrated phonon occupancy of the acoustic mode.  Note that we also used this technique to calibrate the ringdown measurement of Fig.~\ref{fig3}c in units of phonon number.


%

\vspace{2mm}
\noindent\textbf{Data Availability}\\ 
The data that support the findings of this study are available from the corresponding author (OP) upon reasonable request.




\vspace{2mm}
\noindent\textbf{Acknowledgements}\\ 
The authors would like to thank A. Sipahigil for valuable discussions. This work was supported by the AFOSR-MURI Quantum Photonic Matter (FA9550-16-1-0323), the ARO-MURI Quantum Opto-Mechanics with Atoms and Nanostructured Diamond (grant N00014-15-1-2761), the Institute for Quantum Information and Matter, an NSF Physics Frontiers Center (grant PHY-1125565) with support of the Gordon and Betty Moore Foundation, and the Kavli Nanoscience Institute at Caltech. H.R. is supported by the National Science Scholarship from A*STAR, Singapore. 

\vspace{2mm}

\noindent\textbf{Author contributions.}\\ 
HR, GSM, and OP came up with the concept and planned the experiment. HR performed the device design and fabrication. HR, GSM, and MM performed the measurements. HR, MHM, and OP analyzed the data. All authors contributed to the writing of the manuscript.

\vspace{2mm}

\noindent\textbf{Additional information}\\
\textbf{Supplementary information} is available in the online version of the paper.  \\
\textbf{Competing interests.} The authors declare no competing interests.\\
\textbf{Materials \& Correspondence.} Correspondence and requests for materials should be sent to OP (opainter@caltech.edu).

\appendix
\clearpage

\onecolumngrid

\section{Cavity design optimization}
\label{Note1}

\begin{figure}[bp!] 
\begin{center}
\includegraphics[width=0.5\columnwidth]{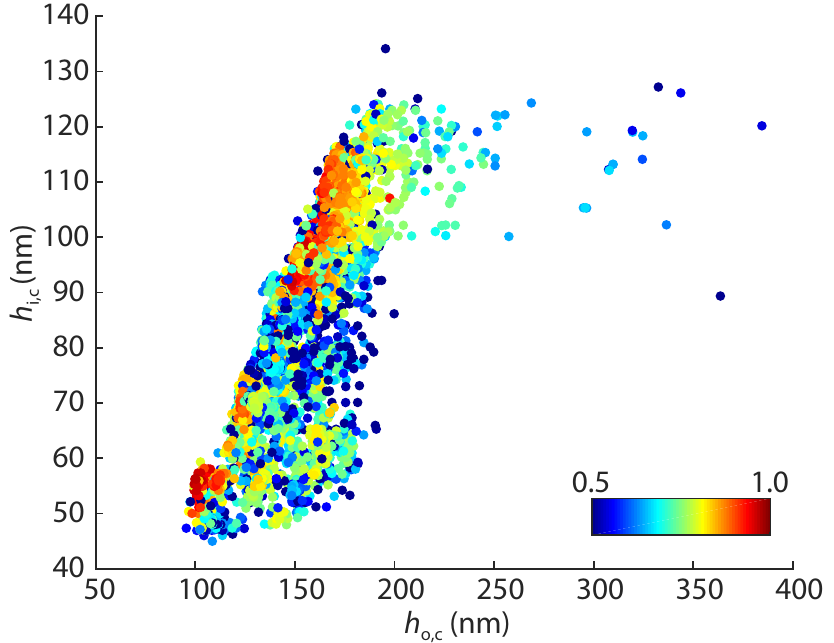}
\caption{\textbf{Nelder-Mead simplex search pattern.} A slice of the multidimensional parameter space explored by the Nelder-Mead minimization method. The color of the points indicate the normalized value of the fitness function. This slice includes multiple Nelder-Mead search runs with randomly generated starting points and convergence to multiple hot-spots in the two-dimensional space of $h_\text{i,c}$ and $h_\text{o,c}$.
}
\label{SI_design_optimization}
\end{center}
\end{figure}

Using a Finite-Element Method (FEM), we simulated the OMC cavity geometry to determine the optical and mechanical cavity mode frequencies ($\omega_\text{o}$ and $\omegam$), vacuum optomechanical coupling rate, $\gzero$, and scattering-limited optical quality factor, $\Qoscat$. 
To maximize $\gammaOM = 4 \gzero^2 \ncav / \kappa$, we attempted to maximize both $\gzero$ and loaded $Q$-factor $\Qo$. Intrinsic quality factors of fabricated devices rarely get higher than $\Qoi = 10^6$ due to fabrication imperfections and optical absorption, while simulated $\Qoscat$ is generally high ($\Qoscat > 5 \times 10^6$) for a properly formed optical cavity. Therefore, we filtered out simulated geometries in which $\Qoscat < 2 \times 10^6$ to prevent radiative scattering from degrading $\Qo$ in the fabricated device. After filtering out low $Q$ designs, we assigned each design a fitness value simply given by $F \equiv - \vert\gzero\vert$. 
We had 9 parameters over which we optimized $F$, which were $d$, $h_\text{i}$, $w_\text{i}$, $h_\text{o}$, $w_\text{o}$, $h_\text{i,c}$, $w_\text{i,c}$, $h_\text{o,c}$, and $w_\text{o,c}$. 
The parameters $h_\text{i,c}$, $w_\text{i,c}$, $h_\text{o,c}$, and $w_\text{o,c}$ are for the `C'-shape holes in the center of the cavity. Parameters of the other cavity `C'-shape holes between the mirror and the center on both sides were vaired quadratically with distance from the center holes. Note $a$, $r$ and $w$ were previously optimized for the large optical and mechanical bandgaps, and the thickness of the device layer, $t=220$~nm, was fixed by the choice of substrate.

For a computationally expensive fitness function with a large parameter space, a good choice of optimization algorithm is the Nelder-Mead method~\cite{Nelder65}. Such a simplex search algorithm 
does not have smoothness requirements for the fitness function, making it quite resistant to simulation noise. A modern variant of this method is also implemented in fminsearch function of MATLAB. An optimization for quasi-2D OMC design is creates as follow: 

\begin{enumerate}

  \item To ensure realizable (ones we can fabricate) geometries are generated in a simulation, the parameter sets need to meet certain conditions. For example, $h_\text{o} - h_\text{i} \geq 60$~nm ($55$~nm for some of iterations) and $w_\text{o}/2 - w_\text{i}/2 \geq 60$~nm, where $60$~nm is a conservative gap size we can realize with the limits of our device fabrication. Therefore, parameter sets are bounded for the generation of initial values and intermediate steps with the Nelder-Mead method.
  \label{opt_step1}
  \item Randomly generate an initial parameter set ($d$, $h_\text{i}$, $w_\text{i}$, $h_\text{o}$, $w_\text{o}$, $h_\text{i,c}$, $w_\text{i,c}$, $h_\text{o,c}$, $w_\text{o,c}$) within the bounds we set in step~\ref{opt_step1}.
  \label{opt_step2}
  \item Run the optical simulation to determine the optical wavelengths ($\omega_\text{o}$) of all the optical modes near 1550 nm with scattering-limited Q-factors larger than a threshold value (in practice only the fundamental mode we are interested in for most cases). If this fails, set $F = 0$ and go to step~\ref{opt_step6}.
  \label{opt_step3}
  \item Scale all parameters except $t$, including $a$, $r$, and $w$, to move the optical mode with highest $\Qoscat$ to approximately 1550 nm.
  \label{opt_step4}
  \item Run the optical simulation again, in addition to the mechanical simulation, with scaled parameters, to determine $\omega_\text{o}$, $\omegam$ and $\gzero$, and compute the fitness of the current scaled parameter set. If $F$ did not change appreciably over the last few iterations, we reached a local
  minimum. Otherwise, we choose a new initial point by going to step~\ref{opt_step6}.
  \label{opt_step5}
  \item Generate a new parameter set via the Nelder-Mead method and go to step~\ref{opt_step3}.
  \label{opt_step6}
  
\end{enumerate} 

By continually repeating the optimization algorithm, we mitigated the problem of converging on a local minimum. A visual representation of the search pattern is shown in supplementary figure~\ref{SI_design_optimization}. We follow these steps until we have a design with a $\gzero$ and $\Qoscat$ that we are satisfied with. 
The visual representation of supplementary figure~\ref{SI_design_optimization} is formed by $\sim 5000$ individual simulations (each individual simulation is defined after step~\ref{opt_step5} has finished successfully). We slice the multidimensional parameter space using $h_\text{i,c}$ and $h_\text{o,c}$ since the gap formed by these two parameters was where both mechanical displacement and optical field are most concentrated. Also, mechanical resonance $\omegam$ highly depends on $h_\text{i,c}$. Indeed, we noticed that most of the local minima lie on the line formed by $h_\text{o,c} - h_\text{i,c} = 60$~nm; this is because the intensity of optical field on the boundaries of the gap becomes stronger as the gap becomes narrower, hence the moving boundary was able contribute a larger amount of coupling due to higher overlap between optical and mechanical fields. This implied that if we could create a narrower gap in the realized devices we could get an even higher $\gzero$. However, for current measurements, we chose a conservative gap value of $h_\text{o,c} - h_\text{i,c} \geq 60$~nm.

\section{Measurement setup}
\label{Note2}

\begin{figure*}[tp!]  
\begin{center}
\includegraphics[width=0.9\columnwidth]{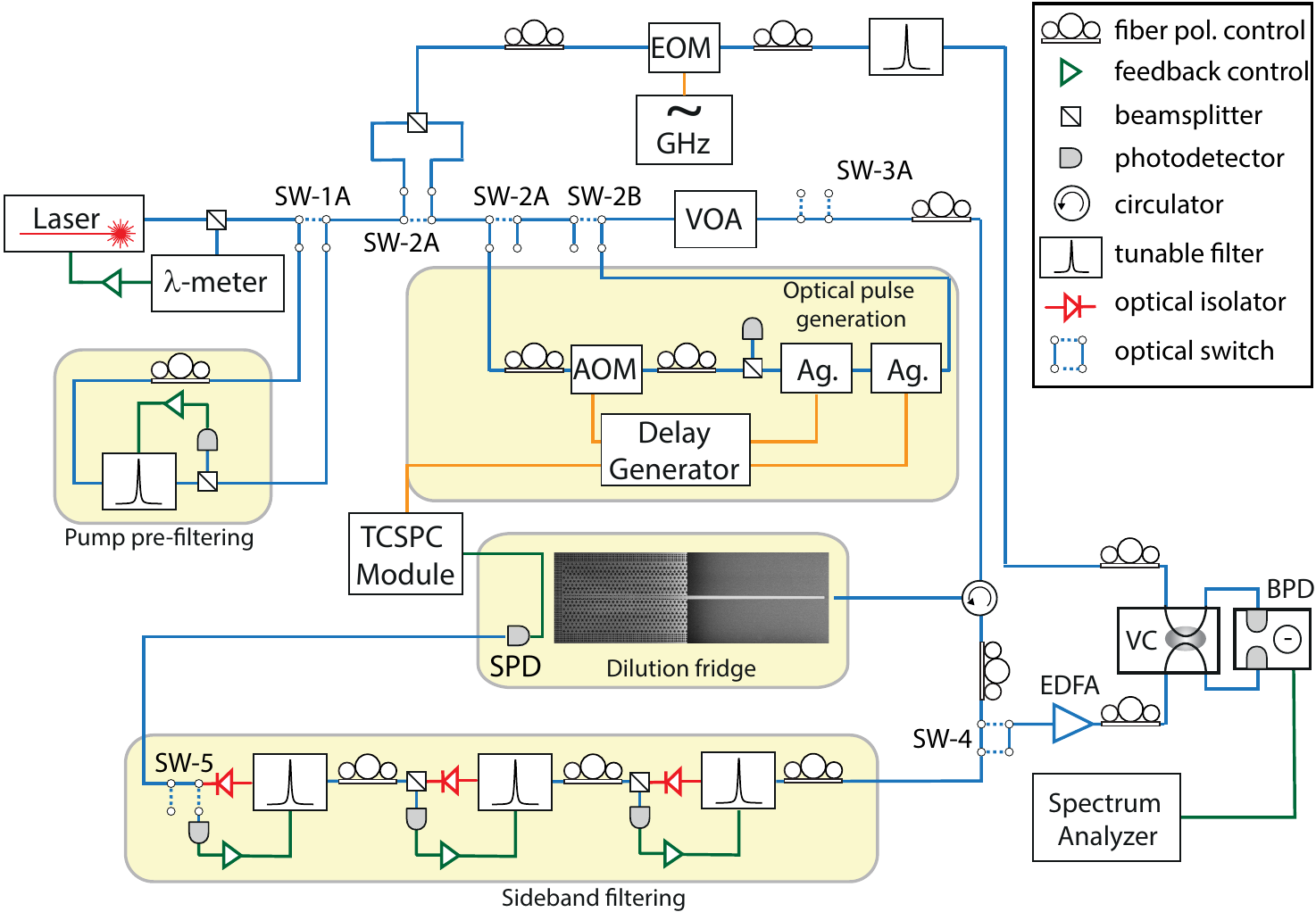}
\caption{\textbf{Diagram of the experimental setup.} A 1550nm external cavity tunable laser is used to generate the optical pump signal in this work.  The laser is initially passed through a single $50$~MHz-bandwidth filter to suppress broadband spontaneous emission noise, after which it can be switched between two different paths: (i) a heterodyne spectroscopy path, and (ii) a photon counting path. In the photon counting path, an acousto-optic modulator (AOM) in series with a fast switch (Ag.) is used for generating high-extinction optical pulses. The modulation components are triggered by a digital delay generator. In the heterodyne spectroscopy path, the light is divided into two paths, one path is passed through an electro-optic intensity modulator (EOM) and a filter to generate the local oscillator (LO) signal, the other path is sent to the optomechanical device. Upon reflection from the device under test, a circulator routes the reflected laser light to either: (i) an EDFA, tunable variable optical coupler (VC), balanced photodiodes (BPD) and spectrum analyzer, or (ii) a sideband-filtering bank consisting of three cascaded fiber Fabry-Perot filters (Micron Optics FFP-TF2) and a SPD operated at $760$~mK. $\lambda$-meter: wavemeter, EOM: electro-optic intensity modulator, AOM: acousto-optic modulator, Ag.: Agiltron 1x1 MEMS switch, SW: optical $2\times2$ switch, VOA: variable optical attenuator, EDFA: erbium-doped fiber amplifier, BPD: balanced photodetector, SPD: single photon detector, TCSPC: time-correlated single photon counting module (PicoQuant PicoHarp 300).} 
\label{SI_setupFig}
\end{center}
\end{figure*}

The measurement setup used for device characterization is shown in Fig.~\ref{SI_setupFig}. A fiber-coupled, wavelength-tunable external cavity diode laser was used as the laser pump in all of our measurements. A small percentage of the laser output was sent to a wavemeter ($\lambda$-meter) for frequency stabilization.
The laser was then passed through an initial $50$~MHz-bandwidth tunable fiber Fabry-Perot filter (Micron Optics FFP-TF2) to reject laser phase noise at the mechanical frequency~\cite{Meenehan2015b}. After this prefiltering, the light could be switched by 2$\times$2 mechanical optical switches between two paths: (i) a balanced heterodyne detection path for performing spectroscopy of the cavity acoustic mode of the quasi-2D OMC cavity, and (ii) a photon counting path in which the laser could be modulated to create a train of high-extinction optical pulses and detection is performed using a single photon detector (SPD) and a time-correlated single photon counting (TCSPC) module. 

For the balanced heterodyne path, a $90:10$ beam-splitter divided the pump laser into local oscillator (LO, 90\%, 0.5 - 1 mW) and signal (10\%) beams. The LO was modulated by an electro-optic modulator (EOM) to generate a sideband at $\delta/2\pi = 50$~MHz above that of the acoustic mode frequency ($\omegam \approx 10$~GHz).  This LO sideband is then selected by high-finesse tunable Fabry-Perot filter before recombining it with the signal. The signal beam is sent to a variable optical attenuator and optical circulator which directs it to a device under test in the dilution refrigerator. The reflected signal beam carrying mechanical noise sidebands at $\omegal \pm \omegam$ was recombined with the LO on a tunable variable optical coupler (VC), the outputs of which were sent to a balanced photodetector (BPD). The detected difference photocurrent contains a beat note corresponding to the acoustic cavity mode response near the LO detuning $\delta$, chosen to lie within the detection bandwidth of the BPD.

For the photon counting path, the laser is prefiltered and routed to an electro-optic phase modulator ($\phi$-EOM) which is driven at the mechanical frequency to generate optical sidebands for locking the pump-cancelling filters (on the detection side). The laser is then directed via 2$\times$2 mechanical optical switches into a `high-extinction' branch consisting of an acousto-optic modulator (AOM, $20$~ns rise and fall time, $50$~dB extinction) and a pair of Agiltron NS 1$\times$1 high-speed switches (Ag., $100$~ns rise time, $30$~$\mu$s fall time, total of $36$~dB extinction).  The AOM and Ag. switches are driven by a digital delay generator to generate high-extinction-ratio optical pulses. The digital delay generator synchronizes the switching of the AOM and Ag. switches with the TCSPC module connected to SPD. 
The total optical extinction of the optical pulses is approximately $86$~dB. After the `high-extinction' branch, the laser light is also passed through the variable optical attenuator (VOA) and a circulator as in the balanced heterodyne path. The reflected signal beam carrying mechanical sidebands was routed to the detection side of the photon counting path. There, the light passes through three cascaded high-finesse tunable fiber Fabry-Perot filters (Micron Optics FFP-TF2) insulated from ambient light, and then sent to a SPD inside the dilution refrigerator.

The SPDs used in this work were amorphous WSi-based superconducting nanowire single-photon detectors developed in collaboration between the Jet Propulsion Laboratory and NIST. The SPDs were designed for a wavelength range $\lambda = 1520 - 1610$~nm, with maximum count rates as large as $10^{7}$ counts per second~(c.p.s.) \cite{Marsili2013}. The SPDs are mounted on the still stage of the dilution refrigerator at $\sim 800$~mK. Single-mode optical fibers are passed into the refrigerator through vacuum feedthroughs and coupled to the SPDs via a fiber sleeve attached to each SPD mount. The radio-frequency output of each SPD is amplified by a cold-amplifier mounted on the $50$~K stage of the dilution refrigerator as well as a room-temperature amplifier, and read out by a triggered PicoQuant PicoHarp 300 time-correlated single photon counting module. After filtering out long-wavelength blackbody radiation inside the DR through a bandpass optical filter and isolating the input optical fiber from environmental light sources at room temperature, we observed SPD dark count rates as low as $\sim0.6$~(c.p.s.) and a SPD quantum efficiency $\etaSPD \simeq 60\%$.\par

The tunable fiber Fabry-Perot filters used for both pre-filtering the pump and filtering the cavity have a bandwidth of $50$~MHz, a free-spectral range of $20$~GHz, and a tuning voltage of $\leq 18$~V per free-spectral range. Each of the filters provides approximately $40$~dB of pump suppression at $10$~GHz offset compared to peak transmission; in total, the filters suppress the pump by $>100$~dB. The three cascaded filter need to be regularly re-locked since they drift during measurement due to both thermal drift and acoustic disturbances in the environment, so they were placed inside a custom-built insulated housing to further improve stability. For the re-locking routine, we switch out of the `high-extinction' (SW-2A,2B) branch and SPD branch (SW-5), as well as the branch which leads to the device under test (SW-3A, the other side connection not shown), to avoid sending large amount of power into the cavity and SPDs. We then drive a EOM (not shown) to generate large optical sidebands on the pump laser signal, one of which is aligned with the cavity resonance ($\omegam/2\pi$).  This modulated signal is sent to the cascaded filters. To re-lock the filter chain, a sinusoidal voltage (0.5 V) was used to dither each filter while monitoring its transmission.
The DC offsets of the dithering sinusoidal signal are then changed while reducing the sinusoidal voltage amplitude to maximize transmission of the desired sideband. After re-locking, the cavity, SPD and `high-extinction' branches are switched back into the optical train, and a new round of measurements can be performed. The total filter transmission was recorded at the end of a re-locking routine and subsequent measurement run, and the previous measurement run was discarded if the transmission shifted by more than a few percent.

\section{Optical coupling to devices}
\label{Note3}

\begin{figure}[tp!] 
\begin{center}
\includegraphics[width=\columnwidth]{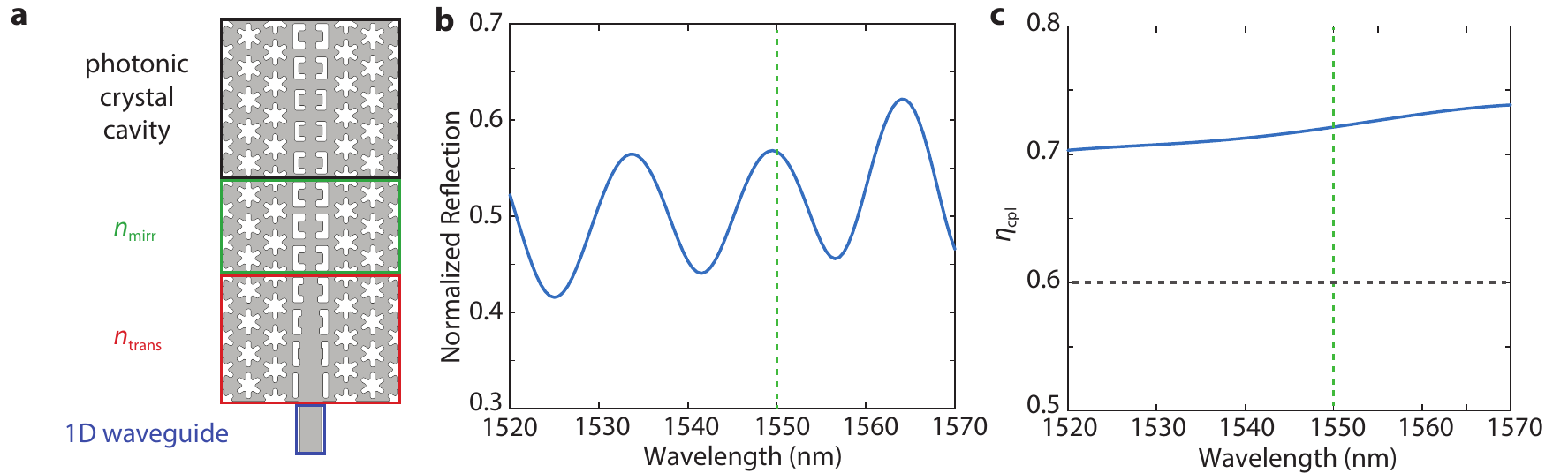}
\caption{\textbf{Optical coupling to devices.} \textbf{a}, Schematic shows the design of the full quasi-2D snowflake OMC device, including central OMC cavity region (black), front mirror section of the OMC cavity (green), OMC cavity to line-defect waveguide transition region (red), and tapered coupling waveguide (blue). \textbf{b}, Broadband reflection spectrum of the optimized coupling waveguide design. \textbf{c}, Plot of the wavelength dependence of the fiber-to-chip coupling inferred from the simulation of the broadband reflection spectrum.  The blue solid curve is the simulated coupling level and the dashed grey curve is the measured single-pass coupling efficiency level, $\eta_\text{cpl}$.  In panel \textbf{b} and \textbf{c} the dashed green vertical line is the nominal operating wavelength.
}
\label{SI_coupler_design}
\end{center}
\end{figure}

The device sample is mounted at the mixing chamber of the DR, with a fiber-to-chip coupling realized by an end-fire coupling scheme with an anti-reflection-coated tapered lensed fiber~\cite{Meenehan2014}. The tapered lensed fiber was placed on a position-encoded piezo xyz-stage in close proximity to the device chip. After cooling the experiment from room temperature to $\sim 10$~mK, we optimize the fiber tip position relative to a tapered 1D coupling waveguide on the device layer by monitoring the reflected optical power on a slow photodetector. \par
The design of the tapered 1D coupling waveguide is similar to those found in Refs.~\cite{Cohen2013} and \cite{Meenehan2014}. The tip of the waveguide was designed to mode match the field of the waist of the lensed fiber. The major distinction for the 2D case was that the other side of the tapered waveguide coupler was also designed to mode match to the line-defect waveguide in the 2D region as shown in Fig.~\ref{SI_coupler_design}a. 
The mirror in the 2D line-defect waveguide region was introduced gradually, to avoid excess scattering in this region. The shape of the center of the line-defect waveguide was slowly changed from a geometry that provides no photonic bandgap, over a number of periods $n_\text{trans}$, to the `C' shape which provides a photonic bandgap. 
Following $n_\text{trans}$, there were a variable number of mirror periods $n_\text{mirr}$.
Reducing $n_\text{mirr}$ made a partially transparent mirror which serves as one side of the cavity's end-mirrors. Thus, a controllable amount of the incident light was permitted to leak through to the cavity region while both of the mirror and defect region of the cavity were highly reflective at frequencies far from resonance.

Supplementary Figure \ref{SI_coupler_design}b shows the broadband reflection spectrum of the optimized coupler, calculated by a finite-difference-time-domain simulation~\cite{Lumerical}. The amplitude and free spectral range of fringes in the spectrum are consistent with a low finesse Fabry-P\'{e}rot cavity formed by weak waveguide-air-interface reflection $R \approx 1.6\%$ and the near unity reflectivity of the quasi-2D OMC cavity mirror. A single-pass coupling efficiency $\eta_\text{cpl}$ is estimated from the broadband reflection spectrum shown in Fig.\ref{SI_coupler_design}c. The actual measured single-pass efficiencies were $\eta_\text{cpl} \approx 60.7\%$ for a zero-shield device and $\eta_\text{cpl} \approx 59.7\%$ for an eight-shield device. The difference between simulations and measurements is attributed to slight fabrication offsets---a small offset on the scale of several nanometers for the width of tapered 1D coupling waveguide may cause significant mode mismatch on both sides.

\section{Acoustic shield}
\label{Note4}

\begin{figure}[tp!]
\begin{center}
\includegraphics[width=0.5\columnwidth]{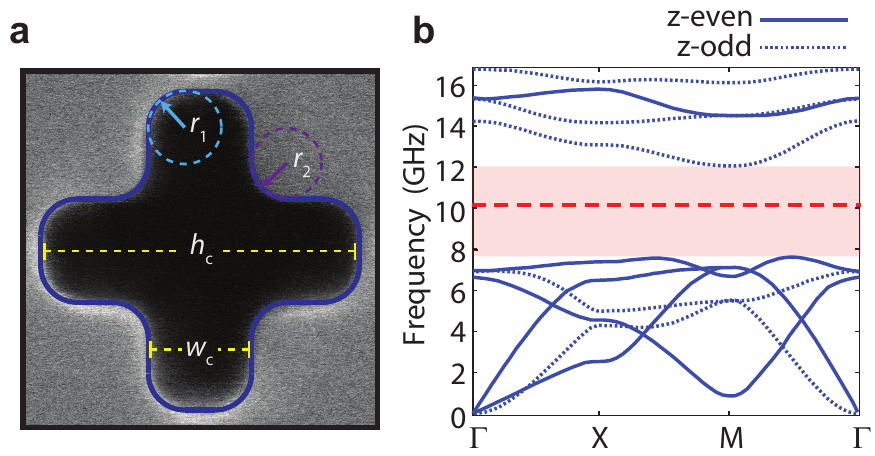}
\caption{\textbf{Cross-structure acoustic bandgap shield.} \textbf{a}, SEM image of an individual unit cell of the cross-crystal acoustic shield. The dashed lines show fitted geometric parameters used in simulation, including cross height ($h_\text{c} = 223$~nm), cross width ($w_\text{c} = 75$~nm), inner fillet radius ($r_\text{1} = 35$~nm), and outer fillet radius ($r_\text{2} = 35$~nm). Thickness of silicon device layer is $220$~nm. 
\textbf{b}, Bandstructure of the realized cross-crystal shield unit cell, with the full  bandgap highlighted in pink. Solid (dotted) lines correspond to modes of even (odd) symmetry in the direction normal to the plane of the unit cell. The dashed red line indicates the mechanical breathing-mode frequency at $\omegam / 2\pi = 10.27$~GHz.}
\label{SI_accoustic_shield_fig}
\end{center}
\end{figure}

To minimize mechanical clamping losses, the quasi-2D OMC was surrounded by an additional shield structure designed to have a complete phononic bandgap at the quasi-2D OMC cavity mode frequency~\cite{Safavi-Naeini2010b, Chan2012}. Geometrically, the structure consists of a square lattice of cross-shaped holes, or equivalently, an array of squares connected to each other via narrow bridges.  We call this the `cross-shield' or `cross-structure'. The phononic bandgap in the cross-structure comes from the frequency separation between the normal-mode resonances of the individual squares and the lower frequency `acoustic bands'.  These acoustic bands are strongly dependent on the width of the connecting narrow bridges, $a_\text{c} \text{-} h_\text{c}$, where $a_\text{c}$ is lattice constant and $h_c$ is the height of cross holes as indicated in Fig~\ref{SI_accoustic_shield_fig}a. 

We analyzed SEM images of fabricated structures to provide parameters for our FEM simulation. For exampe, we included filleting of the inner and outer corners ($r_\text{1}$ and $r_\text{2}$ in Fig.~\ref{SI_accoustic_shield_fig}a) in our simulation, arising from the technical limitations of our nanofabrication methods. The silicon device layer used in the simulations is $220$~nm in thickness, with mass density of $2329~\text{kg}/\text{m}^2$, and anisotropic elasticity matrix ($C_{11}, C_{12}, C_{44}$) = ($166,64,80$)~GPa, assuming a [100] crystallographic orientation along the \textit{x}-axis.
As shown in Fig.~\ref{SI_accoustic_shield_fig}b, a bandgap $> 4$~GHz centered around $\sim 10$~GHz is realized through tuning of the cross lattice constant $a_\text{c}$, cross height $h_\text{c}$ and width $w_\text{c}$.

\section{Thermal conductance simulations of 1D and 2D OMC cavities}
\label{Note5}

\begin{figure*}[tp!]
\begin{center}
\includegraphics[width=\columnwidth]{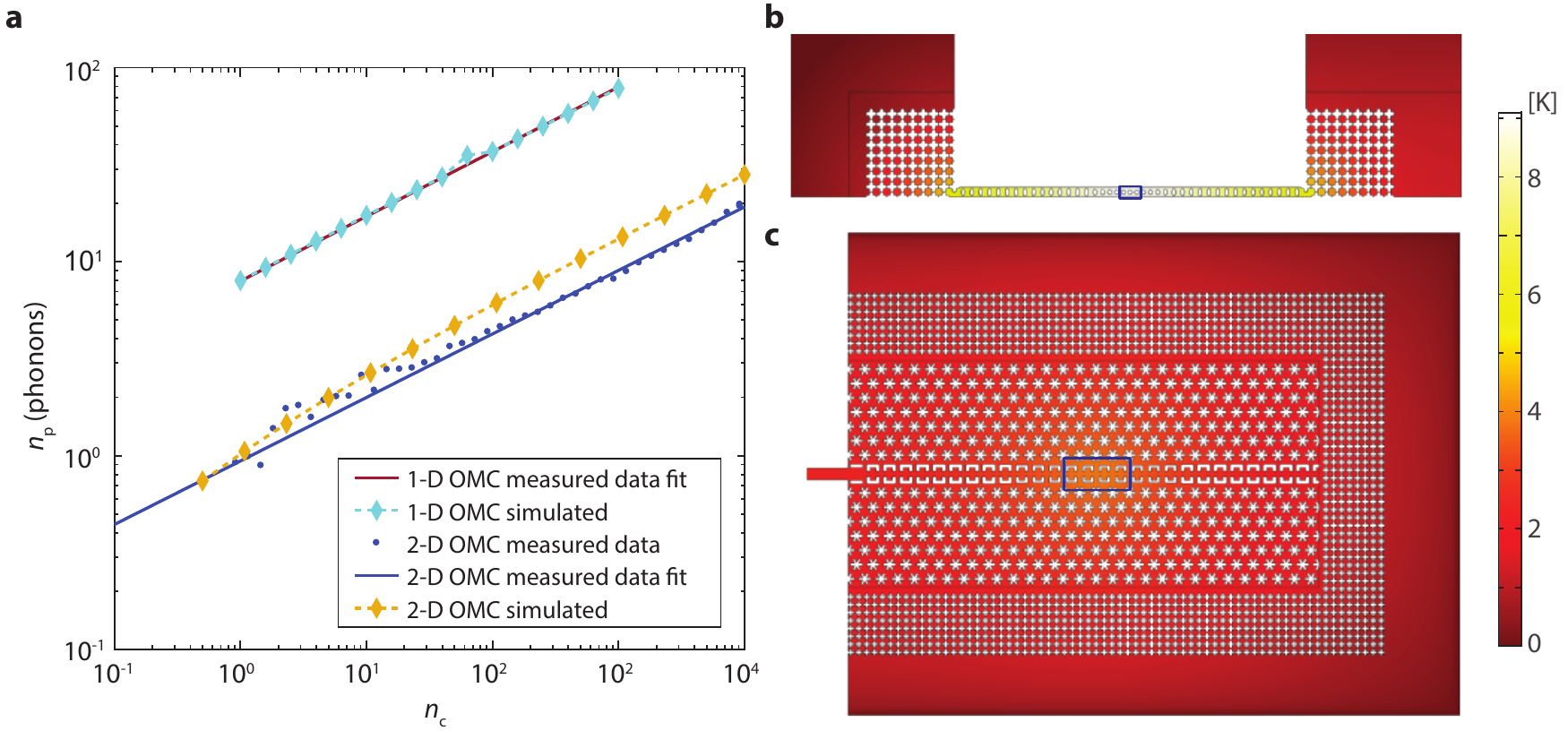}
\caption{\textbf{FEM modelling of thermal conductance.} \textbf{a}, FEM-simulated, measured and fitted curves of $\nbathp$ for both a 1D nanobeam OMC cavity and a quasi-2D OMC cavity versus number of intra-cavity photons $\ncav$. Dashed lines are simulated data and solid lines are fitted curves to measured data. To obtain the material properties used in the simulation, we fit the 1D nanobeam measurement data to a phenomenological thermal conductance model described in the text.  The material properties determined from the simulation of the 1D nanobeam were then used to simulate $\nbathp$ versus $\ncav$ for the quasi-2D OMC cavity. \textbf{b}, FEM-simulated temperature profile of the 1D nanobeam OMC cavity.  \textbf{c}, FEM-simulated temperature profile of the quasi-2D OMC cavity.  In both $\textbf{b}$ and $\textbf{c}$ the intra-cavity photon number is $\ncav=100$ and the temperature scalebar is plotted on the right. The area indicated by blue boxes in the center of both OMC cavities are the heat source used in FEM simulations, where the size the boxes are on the order of optical volume of cavity mode, and total heating power within the boxes volume is $\Pthm$.
The size of the geometries in \textbf{b} and \textbf{c} are not shown on the same scale.} 
\label{SI_thermal_cond}
\end{center}
\end{figure*}

Here we utilize FEM simulations to model the impact of geometry on the thermal conductance of different OMC cavities at millikelvin temperatures.  Specifically our approach is as follows.  We take previous measurements of a 1D nanobeam OMC cavity and compare it to the results of simulations of a similar 1D OMC cavity geometry with variable material properties.  We then find what scaling of the material properties allows us to match experiment to simulation for the 1D OMC cavity.  Using these same scaled values of the material properties we then perform simulations of the quasi-2D OMC cavity. Closing the loop, we find that the simulated values of the hot bath temperature are in correspondence with the measured values for the quasi-2D cavity.  This would indicate that a simple geometric difference in the connectivity of the 1D and 2D cavities to the external chip bath can explain the lower measured hot bath occupancy for the quasi-2D OMC cavity, validating our original design concept.

Under steady state conditions, the power flow from the hot bath into the DR bath, $\Pthm$, is equal to the power flow into the hot bath due to optical absorption.  Here, we have implicitly assumed no other sources of heating other than optical absorption and that the hot bath loses energy via coupling to phonons which radiate into the chip bath at the periphery of the device.  Also assuming the optical absorption process is linear, we find that the power flow into the hot bath is a fraction $\eta_\text{abs}$ of the total input optical power, 
such that $\Pthm = \eta_\text{abs} \Pin \propto \ncav$ (we ignore $\nwg$ here for simplicity). 

For the temperature range considered in this work (where phonon transport is ballistic), the lattice thermal conductivity scales as a power law of the phonon bath temperature~\citep{Holland1963,Callaway1959}. We thus define the thermal conductance from the center of OMC cavity to the DR bath of both the 1D ($C_\text{th,1D}$) and quasi-2D geometries ($C_\text{th,2D}$) such that $\thermcond \propto (\Tbathp)^\alpha$. The exponent $\alpha$ is equal to the effective number of spatial dimensions $d$ of the geometry. The hot bath is assumed to thermalize at an effective temperature $\Tbathp$ and to radiate energy (lattice phonons) into the periphery of the cavity as a black body such that the power lost out of the hot bath goes as $(\Tbathp)^{\alpha+1}$. We can thus write a simple model for the thermal conductance between the hot bath and the periphery of the cavity ($\Tbathnaught$),

\begin{equation}
\begin{aligned}
\Pthm = \thermcond \Delta T \approx \thermcond \Tbathp,\\
\end{aligned}
\label{eqn:thermalconductance}
\end{equation}

\noindent where $\Delta T = \Tbathp - \Tbathnaught$ and in the range of measured $\nbathp$ ($\nbathp > 0.5, \Tbathp > 400$~mK) $\Delta T \approx \Tbathp$ since $\Tbathnaught \ll \Tbathp$. We define $\thermcond$ as $\thermcond = \epsilon (\Tbathp)^\alpha$, where $\epsilon$ depends on the geometry of the cavity and its material properties, which allows us to write,

\begin{equation}
\Pthm = \epsilon \Tbathp^{\alpha+1}=\eta_\text{abs} \Pin \propto \ncav.
\label{eqn:thermalconductancecompare}
\end{equation}

\noindent The power law exponent $\alpha$ in the thermal conductance model is estimated to be $\alpha_0 \approx 2.3$ from the measured data (see Supplementary Figure~\ref{SI_thermal_cond}a).  This is consistent with a Si slab of thickness $t=220$~nm that has an approximately 2D phonon density of states for acoustic modes of frequency in the vicinity of the upper band-edge of the phononic bandgap of the quasi-2D snowflake structure ($\omega/2\pi \gtrsim 10$~GHz).  

Assuming a thermal conductivity for the Si slab which is proportional to $(\Tbathp)^{\alpha_0}$, FEM simulations were performed on both the 1D nanobeam and the quasi-2D snowflake OMC cavity geometries.  As a thermal excitation source we placed a heating source in the center of both OMC cavities with size corresponding to that of the optical mode volume of the cavity mode. The average temperatures within the optical mode volume ($T_\text{p,(1,2)D}$) was then calculated versus intra-cavity photons ($\ncav$) for the 1D nanobeam cavity. We adjusted the material properties (thermal conductivity and absorption coefficient) in order to match the simulated curve to the measured data of the 1D nanobeam cavity from Ref.~\cite{MacCabe2019}.  Finally, we used these adjusted material properties to simulate the quasi-2D cavity. All measured and simulated curves are plotted in Fig.~\ref{SI_thermal_cond}a. We also plot the temperature profile of the 1D and quasi-2D OMC cavities at $\ncav = 100$ in Figs.~\ref{SI_thermal_cond}b and \ref{SI_thermal_cond}c, respectively.

By comparing the simulated curves in Supplementary Figure~\ref{SI_thermal_cond}a, we estimate that the thermal conductance of the quasi-2D and 1D structure has a ratio of $\epsilontwoD/\epsilononeD \approx 42$.  For the same optical pump power applied to the 1D and quasi-2D OMC cavities we have that ${\ncav}_\text{,1D} = {\ncav}_\text{,2D}$ and $\PthoneD = \PthtwoD$.  This yields the relation between thermal conductance and acoustic mode occupancy for the two cavity geometries,

\begin{equation}
\epsilononeD \left(\frac{\hbar{\omegam}_\text{,1D}n_\text{p,1D}}{k_B}\right)^{\alpha_0+1} = \epsilontwoD \left(\frac{\hbar{\omegam}_\text{,2D}n_\text{p,2D}}{k_B}\right)^{\alpha_0+1},
\label{eqn:thermalconductance_occupancy_relation}
\end{equation}

\noindent where we have assumed $\nbathp \approx k_B\Tbathp/\hbar\omegam$ in rewriting the bath temperatures in each cavity in terms of the bath occupancy at the acoustic cavity mode frequency.  Considering that the acoustic mode of the quasi-2D OMC is at frequency ${\omegam}_\text{,2D}/2\pi \approx 10.27$~GHz while that of the 1D resonator is at half this frequency at ${\omegam}_\text{,1D} / 2\pi \approx 5$~GHz, we can write for the ratio of the effective bath occupancies in the two cavities,

\begin{equation}
\frac{n_\text{p,1D}}{n_\text{p,2D}} \approx 2\left(\frac{\epsilontwoD}{\epsilononeD}\right)^{1/(\alpha_0+1)} = 6.2.
\label{eqn:thermalconductance_occupancy_relation_v2}
\end{equation}

\noindent This simulated ratio is in good agreement with the measured ratio of the phonon bath occupancy of the 1D nanobeam OMC cavity in Ref.~\cite{MacCabe2019} and the quasi-2D OMC cavity of this work,  $n_\text{p,1D}/n_\text{p,2D} =7.94/1.1 \approx 7.2$.

\section{Modeling of additional heating in the coupling waveguide}
\label{Note6}

\begin{figure*}[tp!]
\begin{center}
\includegraphics[width=0.7\columnwidth]{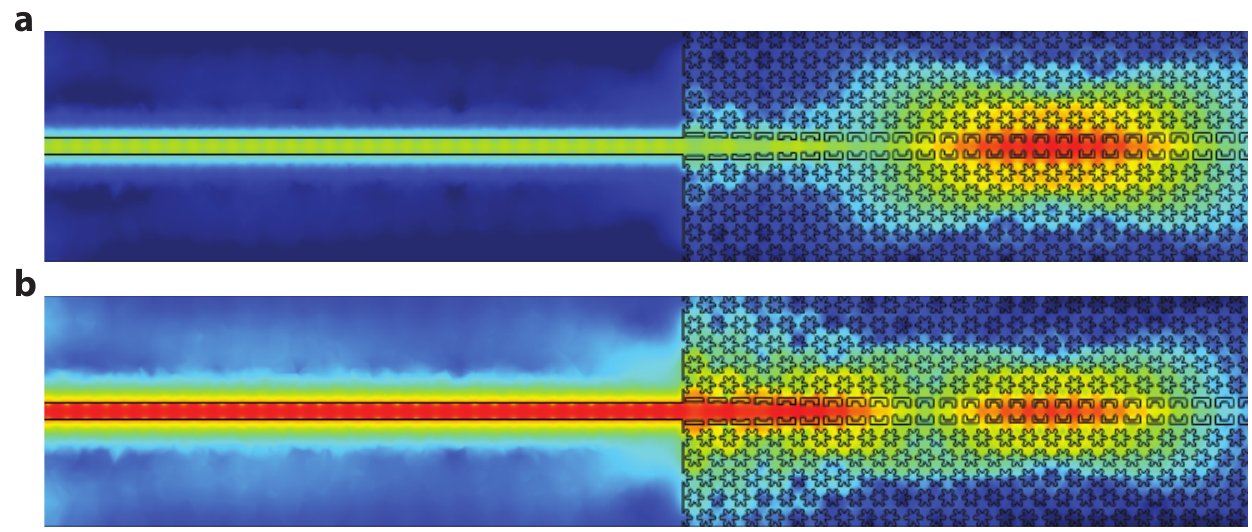}
\caption{\textbf{Simulated optical energy density.} Plot of the FEM-simulated time averaged optical field energy density of the quasi-2D OMC cavity with coupling waveguide for \textbf{a}, $\Delta = 0$~GHz and \textbf{b}, $\Delta = 10$~GHz. Both plots are plotted in logarithm scale and normalized to maximum energy density in each simulation.  
} 
\label{SI_nwg_Sim}
\end{center}
\end{figure*}

In order to better understand the source of the modifies back-action cooling curves measured for the quasi-2D OMC cavities of this work, we performed optical FEM simulations on the full device, including the OMC cavity, 1D coupling waveguide and 2D line-defect waveguide. 
As mentioned, the coupling waveguide in quasi-2D devices was designed to be physically connected to one end of the OMC cavity instead of evanescently coupled to the OMC cavity as in 1D nanobeam OMC devices~\cite{Meenehan2015a,MacCabe2019}. We find below that due to the weak reflectivity of the air-waveguide interface, a weak cavity is formed in the waveguides. As a result, there are two major areas of optical absorption found to be contributing to the hot bath: (i) intra-cavity photons $\ncav$ coupled into the OMC cavity and (ii) photons being coupled into the weak cavity. 
In such a scenario, the occupation of the hot bath $\nbathp$ can depend on both intra-cavity photon number $\ncav$ and the input laser power $\Pin$ in the coupling waveguide.  Here we use an effective waveguide phonon number $\nwg$ ($\nwg$ $\propto$ $\Pin$) to represent the contribution from photons in the weak cavity of the coupling waveguide.

In order to estimate the effect of the optical absorption in the coupling waveguide in comparison to that in the cavity we performed FEM simulations using a geometry which was tuned to approximately the same optical properties as the eight-shield device used for characterizing the hot bath in the main text ($\kappa/2\pi = 1.187$~GHz, $\kappai = 1.006$~GHz).
For optical laser detuning $\Delta = 0$, shown in Fig.~\ref{SI_nwg_Sim}a, a large portion of the input photons that couple into the coupling waveguide are coupled into OMC cavity, with the electric field energy in the OMC cavity one order-of-magnitude higher than in the weak cavity region ($\ncav \gg \nwg$). For optical laser detuning of $\Delta =10$~GHz $\approx \omegam$, shown in Figure~\ref{SI_nwg_Sim}b, a much smaller portion of photons coupled into the coupling waveguides are eventually coupled into OMC cavity due to the large cavity reflection due to the cavity detuning ($\kappa \ll \omegam$). In this case the electric field energy in the OMC cavity is only a few percent of the energy in the weak cavity region ($\ncav \ll \nwg$). Taken together, these two simulations provide strong evidence for the conclusion in the main text that: (i) for on-resonance optical pumping one can ignore the effects of $\nwg$, and (ii) for back-action cooling with laser detuning $\Delta=\omegam$, the optical absorption in the coupling waveguide directly proportional to $\Pin$ should add significant heating of the acoustic mode of the quasi-2D OMC cavity.       

\section{Mode thermalization measurements}
\label{Note7}

\begin{figure*}[tp!]
\begin{center}
\includegraphics[width=0.6\columnwidth]{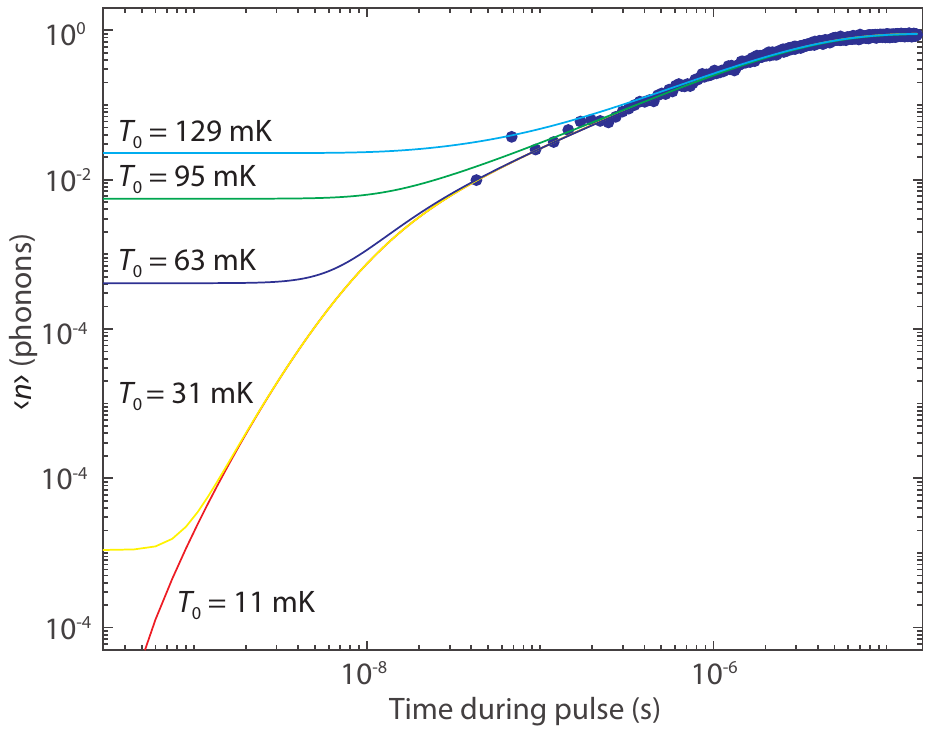}
\caption{\textbf{Acoustic mode occupancy at base temperature.} Plot of the measured (filled blue circles) occupancy of the quasi-2D OMC phonon mode at an applied DR temperature of $\Tf \sim 10$~mK as a function of time within the read-out pulse. Here the read-out photon number is chosen to be small ($\ncav = 9.9$) to minimize parasitic heating during the initial time bins of the pulse. Other measurement parameters are $\Tpulse = 10$~$\mu$s, $\Toff$ = $240$~$\mu$s, and measurement photon-counting bin size $\tau_{\text{bin}}=25.6$~ns. This measurement is performed on a zero-shield device with parameters ($\kappa$, $\kappae$, $\gzero$, $\omegam$, $\gammanotO$) $=$ $2\pi$($1.11$~GHz, $455$~MHz, $1.18$~MHz, $10.238$~GHz, $21.8$~kHz).  Calculated curves based upon a phenomenological heating model~\cite{Meenehan2015b} are shown for $\Tbathnaught = 11$~mK (red), $31$~mK (yellow), $63$~mK (blue), $95$~mK (green), $129$~mK (cyan) are also plotted for reference. 
} 
\label{SI_2DFO_base_occupancy}
\end{center}
\end{figure*}

To measure the true base temperature $\Tbathnaught$ of the quasi-2D OMC cavity devices on the chip we used a low-power ($\ncav=9.9$) laser pump and studied a device with relatively high mechanical damping $\gammanotO$ $=$ $21.8$~kHz (zero-shield device, $Q_\text{m} = 4.69 \times 10^5$) so that data integration time was minimized. With relatively high mechanical damping, the mechanical mode quickly thermalizes to its base temperature between subsequent incident optical pulses so that we could use a rapid measurement repetition rate $1/\Tper$ ($\Tper = \Tpulse + \Toff \gg \gammanotO^{-1}$). The initial mode occupancy during the pulse then approximately corresponds to the base bath occupancy $\nnaught$. In order to remove initial heating from the optical pulse, we fit the entire curve of the measured phonon occupancy throughout the pulse using a phenomenological model of the dynamics of the hot bath and the heating and damping of the acoustic mode~\cite{MacCabe2019}, and extrapolated the fit back to the start of pulse to estimate $\nnaught$.
\par
Supplementary Fig.\ref{SI_2DFO_base_occupancy} shows the measured acoustic mode occupancy versus time within the optical pulse of the quasi-2D OMC cavity.   Theoretical curves from the phenomenological model for $\Tbathnaught = 11$~mK (red), $31$~mK (yellow), $63$~mK (blue), $95$~mK (green), $129$~mK (cyan) are plotted for reference. From these curves the base temperature of the surface of the Si chip is estimated to be $\Tbathnaught \lesssim 63$~mK, corresponding to a base mode occupation of $\nnaught \lesssim 4 \times 10^{-4}$.

\end{document}